\documentclass[twocolumn,aps,showpacs,amssymb,floatfix]{revtex4}
\newcommand{\be}{\begin{equation}}
\newcommand{\ee}{\end{equation}}
\newcommand{\beq}{\begin{eqnarray}}
\newcommand{\eeq}{\end{eqnarray}}
\usepackage{bm}
\usepackage{graphicx}%
\usepackage{epsf}
\usepackage{epsfig}

\usepackage{amsmath}
\usepackage{amsfonts,amssymb}

\begin{document}

\title{Axial Nucleon and Nucleon to $\Delta$ form factors and the Goldberger-Treiman relations  from lattice QCD}
\author{C.~Alexandrou~$^a$, G. Koutsou,$^a$, Th. Leontiou~$^a$, J. W. Negele~$^b$, and A. Tsapalis~$^c$}
\affiliation{{$^a$ Department of Physics, University of Cyprus, CY-1678 Nicosia, Cyprus}\\
{$^b$
Center for Theoretical Physics, Laboratory for
 Nuclear Science and Department of Physics, Massachusetts Institute of
Technology, Cambridge, Massachusetts 02139, U.S.A.}\\{$^c$
Institute for Accelerating Systems and Applications, University of Athens,
Athens,Greece.}}

\date{\today}%

\newcommand{\twopt}[5]{\langle G_{#1}^{#2}(#3;\mathbf{#4};\Gamma_{#5})\rangle}
\newcommand{\threept}[7]{\langle G_{#1}^{#2}(#3,#4;\mathbf{#5},\mathbf{#6};\Gamma_{#7})\rangle}
\newcommand{\GPNN}{$G_{\pi NN}(Q^2)$}
\newcommand{\GPND}{$G_{\pi N\Delta}(Q^2)$}
\newcommand{\GA}{$G_A(Q^2)$}
\newcommand{\GP}{$G_p(Q^2)$}

 \begin{abstract}

 We evaluate the nucleon axial form factor, $G_A(q^2)$, and induced
pseudoscalar form factor,
$G_p(q^2)$, as well as
  the pion-nucleon form factor, $G_{\pi N N}(q^2)$, 
in lattice QCD.
We also evaluate the corresponding nucleon to $\Delta$ transition form 
factors, $C_5^A(q^2)$ and $C_6^A(q^2)$, and the pion-nucleon-$\Delta$
form factor $G_{\pi N\Delta}(q^2)$.  
The nucleon form factors are evaluated 
in the quenched theory and with two degenerate flavors of dynamical
Wilson fermions. The nucleon  to $\Delta$ form factors,
 besides  Wilson fermions,  are evaluated using domain wall valence
fermions with  staggered sea quark configurations
for pion masses as low as about 350 MeV.  Using these
form factors, together with an  evaluation of the
renormalized quark mass, we investigate the validity of the diagonal and
non-diagonal Goldberger-Treiman relations.  
The ratios $G_{\pi N\Delta}(q^2)/G_{\pi NN}(q^2)$ and
$2C_5^A(q^2)/G_A(q^2)$
are constant as a function of the momentum 
transfer squared and show almost no dependence on the quark mass.
We confirm equality of these two ratios consistent with
the Goldberger-Treiman relations extracting a mean value of $1.61(2)$.

\end{abstract}
\maketitle

\section{Introduction}
Form factors measured 
in electromagnetic and weak processes are   fundamental probes of  hadron structure. Despite the long history
of experimental~\cite{exp. nucleon} and theoretical 
studies\cite{theory nucleon}
 on nucleon  
electromagnetic form factors, new measurements of these
quantities continue to reveal
 interesting features. 
The discrepancy between the ratio of the electric to
magnetic nucleon form factors extracted via Rosenbluth separation and
from recent polarization measurements is a well known example.
The transition form factors
in $\gamma N\rightarrow \Delta$ have recently been measured~\cite{bates, jlab}
to high accuracy, paving the way for
theoretical studies using chiral effective
theories~\cite{MarcND, HemmertND} 
and  lattice QCD~\cite{ND lattice, PRL_quenched,
ND dynamical}.
Compared to the electromagnetic form factors,
the nucleon (N) and nucleon  to $\Delta$ form factors 
connected to the axial-vector current  are more difficult to measure and 
therefore less accurately known.
An exception is the  nucleon axial charge  $g_A=G_A(0)$, 
which  can be determined precisely from $\beta-$decay.
Its $q^2$-dependence has been studied 
from neutrino scattering~\cite{Ahrens} 
or pion electroproduction~\cite{Bernard,Choi}. On the other hand
the  nucleon induced pseudoscalar
form factor, $G_p(q^2)$, is less well known.  Muon capture and
radiative muon capture are the main experimental sources
 of information~\cite{Gp review}. 
Both $G_A(q^2)$ and $G_p(q^2)$ have been discussed
within chiral effective theories~\cite{Ga and Gp review, HBCHT}.
The electroweak N to $\Delta$ transition form factors are  even
 less studied. 
 Using Adler's parametrization~\cite{Adler} 
the N to $\Delta$ matrix
element of the axial vector current can be
written in terms of four form factors, 
two of which are
suppressed~\cite{Nimai}.
The two
dominant transition form factors, $C_5^A(q^2)$ and $C_6^A(q^2)$,
are analogous to $G_A(q^2)$ and $G_p(q^2)$, respectively. 
Neutrino interactions in hydrogen and 
deuterium were studied~\cite{barish} in an effort to extract
  information on these form factors.
Experiments using electroproduction of the $\Delta$ resonance
 are under way~\cite{Jlab weak} to measure 
the parity
violating asymmetry in N to $\Delta$, connected to leading
order to the form factor
$C_5^A(q^2)$. Theoretical input on these form factors is
therefore very timely and important. 

State-of-the-art lattice QCD calculations can yield model independent 
results on these axial form factors,
thereby  providing direct comparison with experiment.
Lattice studies  
reflect the experimental situation regarding our
knowledge of these form factors. There have been several 
recent studies
on the electromagnetic nucleon~\cite{NN lattice, LHPC,QCDSF} and N to 
$\Delta$ 
form factors~\cite{ND lattice, PRL_quenched, ND dynamical,  MarcND, HemmertND}.
 There have also been  
 several lattice evaluations 
of $g_A$~\cite{LHPC_axial, QCDSF_axial, Blum_axial}, but
only very recently  lattice studies have began probing  
 the $q^2$-dependence of the nucleon axial form factors~
\cite{LHPC_lat2006, LHPC07} and the
N to $\Delta$ transition form factors~\cite{PRL_axial}.
A notable exception is an early lattice study on the nucleon
axial form factors carried out in the quenched approximation
for rather heavy pion masses~\cite{Liu}.  

In this work we calculate the nucleon axial form factors using
Wilson fermions in the quenched 
theory and with two degenerate flavors of Wilson fermions~\cite{TchiL,Carsten}.
The lowest pion mass in the case of
dynamical Wilson fermions that we use is  about 380 MeV. 
We also evaluate the pion - nucleon 
($\pi NN$)
form factor $G_{\pi N N}(q^2)$. For the extraction of this form
factor we need the renormalized quark mass, which we calculate via the
axial Ward-Takahashi identity (AWI). 
In addition, we present results on  the dominant
axial N to $\Delta$ transition from factors $C_6^A(q^2)$ and $C_5^A(q^2)$. 
The pion-nucleon-$\Delta$ ($\pi N\Delta$) form factor,
 $G_{\pi N\Delta}(q^2)$, is also computed in an analogous way 
to the evaluation of 
$G_{\pi NN}(q^2)$. 
Like in the case of the nucleon axial form factors, the starting point is 
 an evaluation in the quenched theory
using the standard Wilson action. A quenched calculation
 allows an efficient check of our lattice techniques by enabling
a computation of the relevant quantities on  a large
 lattice  minimizing finite volume effects and  obtaining accurate
results  at small momentum transfers reaching  pion mass, $m_\pi$, down to
about 410 MeV.
In the case of the N to $\Delta$ transition,
the light quark regime is  studied in two ways:
 Besides using  configurations with 
two degenerate flavors of dynamical Wilson fermions 
  we use a hybrid combination of domain wall valence quarks, which have chiral
symmetry on the lattice, and
MILC configurations generated with three flavors of staggered sea
quarks  using the Asqtad improved action~\cite{MILC}.
The effectiveness of this hybrid combination has recently been demonstrated 
in the 
successful precision calculation of the nucleon axial charge,  
$g_A$~\cite{LHPC_axial} as well as in our first evaluation of the N to $\Delta$
axial transition form factors~\cite{PRL_axial, athens}.  
Since Wilson fermions
have discretization errors in the lattice spacing, $a$, of ${\cal O}(a)$
 and break chiral symmetry whereas the hybrid action has discretization
 errors of  ${\cal O}(a^2)$ and chirally symmetric valence fermions, 
agreement between calculations using these two
lattice actions provides a non-trivial check of consistency 
of the lattice results.
In this work we obtain results
on the dominant axial N to $\Delta$ form factors $C_5^A$ and $C_6^A$
with improved statistics as compared to their evaluation in
 Refs~\cite{PRL_axial, athens}.

The evaluation of the axial form factors as well as the $\pi NN$ and
$\pi N \Delta$ form
factors allows us to check the Goldberger-Treiman relations.
It is advantageous to calculate 
ratios of the nucleon elastic and transition form factors, since 
ratios have   weak quark mass dependence
and we expect them to be less sensitive 
to other lattice artifacts.
In particular,  it is  useful to consider  ratios
for which the renormalized quark mass cancels since this eliminates
one source of systematic error.

The paper is organized as follows: In Section II, we give the definition
of the matrix elements that we consider in terms of the form factors
on the hadronic level.
In Section III, we give the lattice matrix elements on the 
quark level and in Section IV,
we discuss
how we  extract the form factors
from lattice measurements. In Section V,  we present our results.
Finally in Section VI, we summarize and conclude.

\section{Definition of matrix elements}
To extract the form factors, 
we need to evaluate 
 hadronic matrix elements of the form $<h^\prime|{\cal O}_\mu|h>$,
where $h$ and $h^\prime$  are the initial and final hadron states and
${\cal O}_\mu$ a current that couples to a quark. In all that
follows, we assume isospin symmetry and take the mass of the u and d quark
to be equal. We consider
nucleon-nucleon  and nucleon-$\Delta$ matrix elements of the
axial vector and pseudoscalar currents defined by
\beq
A_{\mu}^a(x)= \bar{\psi}(x)\gamma_\mu \gamma_5\frac{\tau^a}{2}\psi(x) \nonumber\\
P^a(x)= \bar{\psi}(x)\gamma_5 \frac{\tau^a}{2}\psi(x) 
\label{currents}
\eeq
where $\tau^a$ are the three Pauli-matrices acting in flavor space
and $\psi$ the  isospin doublet quark field.

\subsection{Axial form factors}
The matrix element  of the weak axial vector current between nucleon states
can be written in the form
\beq 
\langle N(p',s')|A_\mu^3|N(p,s)\rangle= i \Bigg(\frac{
            m_N^2}{E_N({\bf p}')E_N({\bf p})}\Bigg)^{1/2} \nonumber \\
            \bar{u}(p',s') \Bigg[
            \left(G_A(q^2)\gamma_\mu\gamma_5 
            +\frac{q_\mu}{2m_N}G_p(q^2)\right)\Bigg]\frac{\tau^3}{2}u(p,s)
\label{NN axial}
\eeq 
where we specifically consider the axial isovector current 
 $A^3_\mu$.
The form factors depend only
on the momentum transfer squared, 
$q^2=(p^\prime_\mu-p_\mu)(p^{\prime\mu}-p^\mu)$.
As defined above, the form factors $G_A(q^2)$ and $G_p(q^2)$ are dimensionless.
As already mentioned in the Introduction, there exist several recent lattice studies on the
nucleon axial charge $g_A$ \cite{LHPC_axial, QCDSF_axial, Blum_axial}, 
 whereas only very recently there are lattice studies 
to investigate  the $q^2$ dependence
of  $G_A(q^2)$ or $G_p(q^2)$,  apart from an early calculation in the
quenched approximation~\cite{Liu}.

The invariant N to $\Delta$ weak matrix element is expressed
in terms of four transition
form factors~\cite{Adler,LS} as
\small
\beq
<\Delta(p^{\prime},s^\prime)|A^3_{\mu}|N(p,s)> &=& i\sqrt{\frac{2}{3}} 
\left(\frac{m_\Delta m_N}{E_\Delta({\bf p}^\prime) E_N({\bf p})}\right)^{1/2}
\nonumber \\ 
&\>&\hspace*{-4cm}\bar{u}^\lambda(p^\prime,s^\prime)\biggl[\left (\frac{C^A_3(q^2)}{m_N}\gamma^\nu + \frac{C^A_4(q^2)}{m^2_N}p{^{\prime \nu}}\right)  
\left(g_{\lambda\mu}g_{\rho\nu}-g_{\lambda\rho}g_{\mu\nu}\right)q^\rho \nonumber \\
&\>&\hspace*{-2cm}+C^A_5(q^2) g_{\lambda\mu} +\frac{C^A_6(q^2)}{m^2_N} q_\lambda q_\mu \biggr]u(p,s)
\label{ND axial}
\eeq
\normalsize
where, as in the nucleon case,  we consider the
physically relevant axial isovector current $A^3_\mu(x)$.

\subsection{Pseudoscalar matrix elements}
 Spontaneous symmetry breaking couples pions to the broken axial charges 
and currents. The relation
\be 
\langle 0|A_\mu^a(0)|\pi^b(p)\rangle = i f_\pi p_\mu \delta^{ab}
\label{pion decay}
\ee
can be used to 
extract the pion decay constant,
$f_\pi$, on the lattice by evaluating  two point functions.
With our conventions   $f_\pi= 92$~MeV. Taking the divergence of the
axial vector current we obtain the operator relation
\be  
\partial^\mu A_\mu^a=f_\pi m_\pi^2 \pi^a 
\label{PCAC}
\ee
known as the partially conserved axial vector current (PCAC) hypothesis.
On the QCD level we have the axial Ward-Takahashi identity 
\be 
 \partial^\mu A_\mu^a=2m_qP^a
\label{ward}
\ee
where all quantities appearing in Eq.~(\ref{ward}) are  renormalized
quantities with $m_q$ being the renormalized quark mass.
Comparing Eqs.~(\ref{PCAC}) and (\ref{ward}) 
we can write the pion field $\pi^a$ in terms of the 
pseudoscalar density as
\be
\pi^a=\frac{2m_qP^a}{f_\pi m_\pi^2} \quad.
\label{pion field}
\ee
The renormalized quark mass can be evaluated by taking 
the matrix element of Eq.~(\ref{ward}) between a zero momentum
pion state and
the vacuum to obtain
\be 
m_q = \frac{m_\pi<0|A_0^a|\pi^a(0)>}{2<0|P^a|\pi^a(0)>}
 \quad.
\label{quark mass}
\ee 
 Taking the matrix element of the pseudoscalar density between 
nucleon states we can define the $\pi NN$ form factor via
\beq
 2m_q<N(p^\prime)|P^3|N(p)>= 
\Bigg(\frac{m_N^2}{E_N({\bf p}')E_N({\bf p})}\Bigg)^{1/2} \nonumber \\
\frac{f_\pi m_\pi^2\> G_{\pi NN}(q^2)}
{m_\pi^2-q^2}\>\bar{u}(p^\prime)i\gamma_5 u(p) \quad .
\label{g_piNN}
\eeq
Similarly the nucleon-$\Delta$ matrix element of the pseudoscalar
density yields the $\pi N \Delta$ form factor: 
\beq 
 2m_q<\Delta(p^\prime)|P^3|N(p)> =
\left(\frac{m_\Delta m_N}{E_\Delta({\bf p}^\prime) E_N({\bf p})}\right)^{1/2}
\nonumber \\ 
\sqrt{\frac{2}{3}}\frac{f_\pi m_\pi^2 \>G_{\pi N\Delta}(q^2)}
{m_\pi^2-q^2}
\bar{u}_\nu(p^\prime)\frac{q^\nu}{2m_N} u(p)
\label{g_piND} \quad.
\eeq
Eqs.~(\ref{g_piNN}) and (\ref{g_piND}) define the form factors
  $G_{\pi NN}(q^2)$  and  $G_{\pi N \Delta}(q^2)$ that
we study in this work. The
 $\pi NN$ and $\pi N\Delta$
strong coupling
constants are then
given by $g_{\pi NN}=G_{\pi NN}(0)$ and $g_{\pi N\Delta}=G_{\pi N\Delta}(0)$.
PCAC relates the axial form factors $G_A$ and $G_p$ 
with  $G_{\pi NN}$ 
and equivalently $C_5^A$ and $C_6^A$ with $G_{\pi N\Delta}$.
Using the PCAC hypothesis
together with Eq.~(\ref{g_piNN}) we obtain
the diagonal Goldberger-Treiman relation (GTR)
\beq 
 G_A(q^2)+\frac{q^2}{4m_N^2} G_p(q^2) = 
\frac{1}{2m_N}\frac{2G_{\pi N N}(q^2)f_\pi m_\pi^2}{m_\pi^2-q^2} \quad.  
\label{GTR}
\eeq
Similarly using Eq.~(\ref{g_piND}) we obtain 
the non-diagonal GTR
\beq
 C_5^A(q^2)+\frac{q^2}{m_N^2} C_6^A(q^2) = 
\frac{1}{2m_N}\frac{G_{\pi N \Delta}(q^2)f_\pi m_\pi^2}{m_\pi^2-q^2} \quad.
\label{GTR_ND}
\eeq
Assuming pion pole dominance we can
relate the form factors $G_p$  to $G_{\pi NN}$ and $C_6^A$ to $G_{\pi N\Delta}$
via:
\beq \nonumber
\frac{1}{2m_N}G_p(q^2)&\sim &\frac{2G_{\pi NN}(q^2) f_\pi}{m_\pi^2-q^2} \\
\frac{1}{m_N}C_6^A(q^2)&\sim&\frac{1}{2}\frac{G_{\pi N\Delta}(q^2) f_\pi}
{m_\pi^2-q^2}
\label{GP&CA6}
\eeq
Substituting in Eqs.~(\ref{GTR}) and (\ref{GTR_ND}) we obtain the simplified
Goldberger-Treiman relations
\beq
G_{\pi NN}(q^2)\>f_\pi &=&m_N G_A(q^2) \nonumber \\
G_{\pi N \Delta}(q^2)\>f_\pi &=& 2m_N C_5^A(q^2)
\quad.
\label{GTR simple}
\eeq

\section{Lattice evaluation of correlation functions}

\begin{figure}[h]
\epsfxsize=8truecm
\epsfysize=4truecm
\mbox{\epsfbox{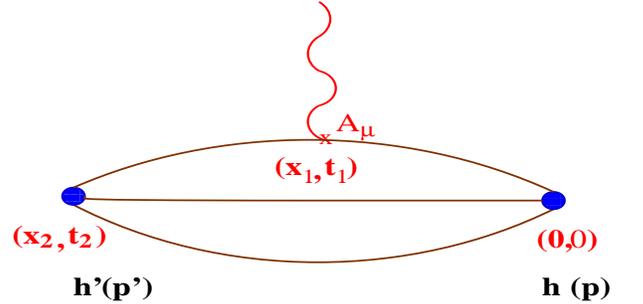}}
\caption{Connected three-point function between final and initial hadron states
 $h^\prime({\bf p}^\prime)$ and $h({\bf p})$.}
\label{fig:3pt diagram}
\end{figure}

To evaluate  the axial nucleon form factors $G_A$ and
$G_p$, we use the techniques developed in
our study of the  nucleon isovector electromagnetic
form factors~\cite{NN lattice}. 
Since only the isovector
axial vector is of relevance here,
only the connected 
diagram  shown in  Fig.~\ref{fig:3pt diagram} is needed.
To extract the matrix element of the  axial isovector current
between  nucleon states defined in Eq.~(\ref{NN axial})
 we need to calculate the three-point function 
\small 
\beq &&
\hspace{-0.3cm} \langle G^{N A_\mu^3 N} (t_2, t_1 ;
{\bf
p}^{\;\prime}, {\bf p}; \Gamma) \rangle  \nonumber \\
 &&= \sum_{{\bf x}_2, \;{\bf
x}_1} \exp(-i {\bf p}^{\;\prime} \cdot {\bf x}_2 ) \exp(+i ({\bf
p}^{\;\prime} -{\bf p}) \cdot {\bf x}_1 ) \; \Gamma^{\beta \alpha}
\nonumber \\ &&\langle \;\Omega \; | \;
T\left[\chi^{\alpha} ({\bf x}_2,t_2) A_\mu^3({\bf
x}_1,t_1) \bar{\chi}^{\beta} ({\bf 0},0) \right] \; | \;\Omega
\;\rangle   \; , 
\label{3pt}
\eeq 
\normalsize 
using the local quark bilinear axial current
$A_{\mu}^3(x)$ of Eq.~(\ref{currents}).
An interpolating field with the quantum numbers of the nucleon
routinely used in lattice studies
is 
\be 
\chi  (x) =
\epsilon^{a b c}\; \left[ u^{T\; a}(x)\; C \gamma_5 d^b(x)
\right]\; u^c(x)  \quad.
\label{nucleon field}
\ee
We create an initial state (source) by acting
with $\bar{\chi}(0)$ on the vacuum. 
Evolution in Euclidean time with the QCD Hamiltonian produces, in the large
time limit, the nucleon state. We take $t_1$, the time from
the source at which the
axial-vector current couples to a quark, to be large enough so that
the nucleon  is the dominant state. We then take the overlap with  a 
nucleon state that is annihilated at a later time
 $t_2-t_1$ by the same interpolating 
field $\chi(x)$ (sink). Again we take $t_2-t_1$ large enough
so that the nucleon is the dominant state.
This process is schematically shown in 
Fig.~\ref{fig:3pt diagram}. 
In addition, we calculate the nucleon  two-point function,
\small 
\beq
\hspace{-0.2cm}  
\langle G^{NN} (t, {\bf p} ; \Gamma) \rangle 
 = \hspace{-0.1cm} \sum_{{\bf x}} e^{-i {\bf p} \cdot {\bf
x} } \; \Gamma^{\beta \alpha}\,\langle \Omega
|\;T\;\chi^{\alpha}({\bf x},t)
 \bar{\chi}^{\beta} ({\bf 0},0)
\; |  \Omega\;\rangle
\eeq 
\normalsize
where the
 projection matrices for the Dirac indices are given by
\be
\Gamma_i = \frac{1}{2}
\left(\begin{array}{cc} \sigma_i & 0 \\ 0 & 0 \end{array}
\right) \;\;, \;\;\;\;
\Gamma_4 = \frac{1}{2}
\left(\begin{array}{cc} I & 0 \\ 0 & 0 \end{array}
\right) \;\; .
\ee
We then construct a ratio such that, in the large Euclidean time limit
all exponential dependences on times and
unknown initial state-nucleon overlap constants 
$<N|\bar{\chi}|0>$ cancel. One can find more than one ratio that
 accomplishes this. We require, in addition, that we use  two
point functions that involve the shortest time evolution. Such a ratio
is given by
\small
\beq
R^{A}(t_2, t_1; {\bf p}^{\; \prime}, {\bf p}\; ; \Gamma ; \mu) &=&
\frac{\langle G^{NA_\mu^3 N}(t_2, t_1 ; {\bf p}^{\;\prime}, {\bf p};\Gamma ) \rangle \;}{\langle G^{N N}(t_2, {\bf p}^{\;\prime};\Gamma_4 ) \rangle \;} \> \nonumber \\
&\>& \hspace*{-3.8cm}\biggl [ \frac{ \langle G^{N N}(t_2-t_1, {\bf p};\Gamma_4 ) \rangle \;\langle 
G^{NN} (t_1, {\bf p}^{\;\prime};\Gamma_4 ) \rangle \;\langle 
G^{N N} (t_2, {\bf p}^{\;\prime};\Gamma_4 ) \rangle \;}
{\langle G^{N N} (t_2-t_1, {\bf p}^{\;\prime};\Gamma_4 ) \rangle \;\langle 
G^{N N} (t_1, {\bf p};\Gamma_4 ) \rangle \;\langle 
G^{N N} (t_2, {\bf p};\Gamma_4 ) \rangle \;} \biggr ]^{1/2} \nonumber \\
&\;&\hspace*{-1cm}\stackrel{t_2 -t_1 \gg 1, t_1 \gg 1}{\Rightarrow}
\Pi^A({\bf p}^{\; \prime}, {\bf p}\; ; \Gamma ; \mu) \; ,
\label{R-ratio}
\eeq
\normalsize
which, in the large Euclidean time, with  the nucleon
the dominant state,  produces a constant  
plateau region in $t_1$. 
Throughout this work we use kinematics where the final hadron state
 is produced at rest
and therefore the momentum transfer 
${\bf q}={\bf p}^{\prime}-{\bf p}=-{\bf p}$.
We take
$-q^2=Q^2>0$
with $Q^2$ being the Euclidean momentum transfer squared.
The value of
$R^A(t_2, t_1; {\bf p}^{\; \prime}, {\bf p}\; ; \Gamma ; \mu)$
in the plateau region, $\Pi^A ({\bf p'}, {\bf p}\; ;\Gamma ;\mu)$, is
directly connected to 
the nucleon form factors
through the relation
\beq  
\Pi^A (0, -{\bf q}\; ;\Gamma_k ;\mu) &=& i \;\frac{C}{4 m_N}\nonumber \\ 
& &\hspace*{-3.8cm} \biggl[ \bigl( (E_N+m_N ) \delta_{k,\mu} +
 q_k  \delta_{\mu,4} \bigr)  G_A(Q^2)
- \frac{q_\mu  q_k}{2 m_N} G_p(Q^2) \biggr] 
\eeq 
for $k = 1,2,3$, while 
$\Pi^A (0, -{\bf q}\; ;\Gamma_4 ;\mu) = 0$.
The nucleon energy $E_N = \sqrt{m_N^2 + {\bf q}^2}$ 
 and $C= \sqrt{\frac{2 m_N^2}{E_N(E_N + m_N)}}$, a factor 
related to the normalization of the lattice states.
Since our goal is to  evaluate
 the form factors as a function of $Q^2$, 
we calculate  the three point
functions with sequential inversions through the sink.
This requires fixing the source-sink time separation  $t_2$  as well
as the initial and
final hadron states but allows the insertion of any operator
with arbitrary momentum and any
time slice $t_1$. 
In fact, the usefulness of this technique is evident  in this work:
Since the same matrix elements were calculated for the electromagnetic 
current~\cite{NN lattice,PRL_quenched, ND dynamical}
{\it no new sequential inversions} are required for the axial current
or pseudoscalar density operators.

As in the electromagnetic case, it is advantageous
to use a linear combination of nucleon interpolating fields 
to construct optimal sources and sinks.
Since for axial operators a non-zero contribution can be obtained 
only if
$\Gamma \neq \Gamma_4$ in the three-point function,  the most symmetric
linear combination of  matrix elements that can be considered is 
\beq  
S^A({\bf q}\; ;\; j)= \sum_{k=1}^3\Pi^A ({\bf 0}, -{\bf q}\; ;
\Gamma_k ;\mu = j) = \nonumber \\ 
i \; \frac{C}{4m_N}\biggl[(E_N+m_N)
\left( \delta_{1,j} + \delta_{2,j} + \delta_{3,j} \right) G_A(Q^2)
\nonumber \\
- (q_1 + q_2 + q_3) \frac{q_j}{2 m_N} G_p(Q^2) \biggr] \quad,
\label{SA optimal}
\eeq 
where $j = 1,2,3$ labels the spatial current direction. 
In order to obtain the three matrix elements corresponding to three
choices of $\Gamma_k$ in the above sum one would require three sequential
inversions. However, 
 choosing an appropriate linear combination of nucleon interpolating fields,
this sum is automatically built in and with one sequential inversion we
can obtain $S^A({\bf q};j)$ for all current directions $j$.
We call such a linear combination optimal sink because it allows us
to take into account in our determination of the
form factors the largest set of  momentum vectors
contributing to the same $Q^2$ value.
Since the sequential propagators corresponding to this sink have  been 
computed for the isovector
electromagnetic form factors~\cite{NN lattice} they  can be used 
directly  here. 
Therefore the computational cost for the evaluation of the three-point
function for all intermediate times $t_1$, current indices $\mu$ and 
a large set
of lattice momenta vectors ${\bf q}$ is very small.

Similarly, to evaluate the $\pi NN$ form factor $G_{\pi NN}$ 
we construct the ratio $R^P$, 
which is the same as the ratio  $R^A$  given in Eq.~(\ref{R-ratio}) 
but instead of the three-point function $\langle G^{N A^3_\mu N} (t_2, t_1 ;
{\bf p}^{\;\prime}, {\bf p}; \Gamma) \rangle$ defined in Eq.~(\ref{3pt}) with
 the  axial current $A^3_\mu$, 
we use the three-point function 
$\langle G^{N P^3 N} (t_2, t_1 ;
{\bf p}^{\;\prime}, {\bf p}; \Gamma) \rangle$,
obtained by replacing $A^3_\mu$ in Eq.~(\ref{3pt})
by  the pseudoscalar density $P^3$.
 The large Euclidean time behavior of $R^P$ is independent
of $t_1$ leading to the plateau value denoted by $\Pi^P ({\bf 0}, -{\bf q}\; ;
\Gamma ;\gamma_5)$.
The value of $R^P$ in the plateau region, $\Pi^P$,  is related to
the $\pi NN$ form factor via
\beq  
\Pi^P (0, -{\bf q}\; ;
\Gamma_k ;\gamma_5) &=& \nonumber \\ 
&\>&\hspace*{-2cm} C \; \frac{q_k}{2 m_N} \frac{f_\pi m_\pi^2} 
{2 m_q (m_\pi^2 + Q^2)} \; G_{\pi NN} (Q^2)  
\eeq 
for $k = 1,2,3$ while 
$\Pi^P (0, -{\bf q}\; ;\Gamma_4 ;\gamma_5) = 0$.
 Summation over the polarized matrix elements now leads to
\beq  
S^P({\bf q}\; ;\; \gamma_5)&=& \sum_{k=1}^3\Pi^P (0, -{\bf q}\; ;
\Gamma_k ;\gamma_5) = \nonumber \\ 
 &\>& \hspace*{-1.8cm}
C\>\frac{q_1 + q_2 + q_3}{2 m_N} \frac{f_\pi m_\pi^2} 
{2 m_q (m_\pi^2 + Q^2)} \; G_{\pi NN} (Q^2)  \quad,
\label{P optimal}
\eeq 
from which $G_{\pi NN}$ can be extracted if we know $f_\pi$ and $m_q$.

The determination of the N to $\Delta$ axial transition form factors
requires the
evaluation of the three-point function
\small 
\beq &&
\hspace{-0.3cm} \langle G^{\Delta A_\mu^3 N}_{\sigma} (t_2, t_1 ;
{\bf
p}^{\;\prime}, {\bf p}; \Gamma) \rangle  \nonumber \\
 &=& \sum_{{\bf x}_2, \;{\bf
x}_1} \exp(-i {\bf p}^{\;\prime} \cdot {\bf x}_2 ) \exp(+i ({\bf
p}^{\;\prime} -{\bf p}) \cdot {\bf x}_1 ) \; \Gamma^{\beta \alpha}
\nonumber \\ &&\langle \;\Omega \; | \;
T\left[\chi^{\alpha}_{\sigma}({\bf x}_2,t_2) A_{\mu}^3({\bf
x}_1,t_1) \bar{\chi}^{\beta} ({\bf 0},0) \right] \; | \;\Omega
\;\rangle   \;  \quad,
\eeq 
\normalsize
where to create an initial state with  the $\Delta^+$ quantum numbers
 we use the standard Rarita-Schwinger 
interpolating field 
\beq 
\chi
^{\Delta^{+}}_\sigma (x) &=&\hspace{-0.2cm} \frac{1}{\sqrt{3}}
\epsilon^{a b c} \Big \lbrace 2 \left[
u^{T a}(x)\; C \gamma_\sigma d^b(x) \right]u^c(x) \; \nonumber \\
&+&\hspace{-0.2cm} \left[ u^{T a}(x)\; C \gamma_\sigma u^b(x)
\right]d^c(x) \Big \rbrace . 
\label{Delta field}
\eeq
Besides the nucleon two-point function
we also need  the $\Delta$ two-point function given by
\small
\beq
&&\hspace*{-1cm} \langle G^{\Delta \Delta}_{\sigma\tau} (t, {\bf p}^{\;\prime} ;
\Gamma) \rangle \nonumber \\
&=&  \sum_{{\bf x}} e^{-i {\bf p}^{\;\prime} \cdot {\bf x} } \;
\Gamma^{\beta \alpha}\;\langle \Omega
|\;T\;\chi^{\alpha}_{\sigma}({\bf x},t)
 \bar{\chi}^{\beta}_{\tau} ({\bf 0},0)
\; | \Omega \;\rangle \quad.
\label{DD} 
\eeq 
\normalsize
The corresponding ratio which, in the large Euclidean time limit 
 becomes $t_1$-independent, is given by 
\small 
\beq 
\label{R2-ratio}
\vspace*{-0.8cm} 
R^{A}_\sigma (t_2, t_1; {\bf p}^{\; \prime}, {\bf p}\; ; \Gamma ; \mu)
&=&   \frac{\langle G^{\Delta A_\mu^3 N}_{\sigma} (t_2,
t_1 ; {\bf p}^{\;\prime}, {\bf p};\Gamma ) \rangle \;} {\langle
G^{\Delta \Delta}_{ii} (t_2, {\bf p}^{\;\prime};\Gamma_4 ) \rangle
\;}   \nonumber \\
&\>& \hspace{-3.9cm} \biggr [\frac{\langle G^{\Delta \Delta}_{ii} (t_2, {\bf
p}^{\;\prime};\Gamma_4 ) \rangle}{ \langle
G^{N N} (t_2, {\bf p};\Gamma_4 ) \rangle }\> 
\hspace*{0.0cm}\frac{ \langle G^{N N}(t_2-t_1, {\bf
p};\Gamma_4 ) \rangle \;\langle G^{\Delta \Delta}_{ii} (t_1, {\bf
p}^{\;\prime};\Gamma_4 ) \rangle} {\langle G^{\Delta \Delta}_{ii}
(t_2-t_1, {\bf p}^{\;\prime};\Gamma_4 ) \rangle \;\langle
G^{N N} (t_1, {\bf p};\Gamma_4 ) \rangle} \biggr ]^{1/2} \nonumber \\
&\;&\hspace*{-1.0cm}\stackrel{t_2 -t_1 \gg 1, t_1 \gg
1}{\Rightarrow} \Pi^A_{\sigma}({\bf p}^{\; \prime}, {\bf p}\; ;
\Gamma ; \mu) \; \quad. 
\label{RND-ratio}
\eeq
\normalsize
As in the nucleon case, we take the final 
$\Delta$ state to be produced 
at rest and therefore ${\bf q}={\bf p}^{\prime}-{\bf p}
=-{\bf p}$.
The value of $R^A_\sigma$ in the plateau region, $\Pi^A_\sigma$,
for the case $\Gamma=\Gamma_4$, which is the relevant one for this work, 
is related to
the form factors via
\beq \Pi_k^A&\>&\hspace*{-0.5cm}
({\bf 0},-\mathbf{q};\Gamma_4;j)=i B\Bigg[
-\bigg\lbrace\frac{(E_N-2m_\Delta+m_N)}{2}\delta_{k,j}\nonumber \\
&\>&+\frac{p^kp^j}{2(E_N+m_N)}\bigg\rbrace C^A_3 \nonumber
-\left\{(E_N-m_\Delta)\frac{m_\Delta}{m_N}\delta_{k,j}\right\}C^A_4\\
&\>&+m_N\delta_{k,j}C^A_5 -\frac{p^kp^j}{m_N}C^A_6 \Bigg] \eeq 
for
the spatial components of the current, $\mu=j$,whereas
for the temporal current component, $\mu=4$, we have
\be
\Pi_k^A({\bf 0},-\mathbf{q};\Gamma_4;4)=B\>p^k\left[C^A_3+\frac{m_\Delta}{m_N}C^A_4+\frac{E_N-m_\Delta}{m_N}C^A_6\right]
\quad, \ee 
for $k=1,2,3$, 
while
$\Pi_4(0,-\mathbf{q};\Gamma;\mu)=0$. 
In this case we have a larger freedom in choosing  appropriate
linear combinations for the optimal sink due to the additional
vector index of the $\Delta$. Using this freedom
we construct  $\Delta$ sinks
so that the maximum allowed number of lattice vectors contribute
in the evaluation of the form factors at a given value of $Q^2$. 
These turn out to be the same as the ones used in our study of the
 N to $\Delta$ 
electromagnetic transition form factors~\cite{PRL_quenched}.
Therefore the sequential inversions already performed for the evaluation of the
N to $\Delta$ electromagnetic transition form 
factors~\cite{PRL_quenched, ND dynamical}
can be  used for the computation of the axial transition form factors.
We give below  the expressions that we obtain
in the large Euclidean
time limit, using
the optimal linear combinations of $\Delta$ interpolating
fields:  
\small
\beq 
 S_1^A(\mathbf{q};j) &= &\sum^3_{\sigma=1}\Pi^A_\sigma({\bf 0},-\mathbf{q};\Gamma_4;j) 
= i B\nonumber \\
&&\hspace{-1.5cm} \Bigg[-\frac{C^A_3}{2}\bigg\lbrace (E_N-2m_\Delta+m_N)+
\left(\sum_{k=1}^3 p^k\right)\frac{p^j}{E_N+m_N} \bigg\rbrace 
\nonumber \\ 
&\>&\hspace{-1.5cm} -\frac{m_\Delta}{m_N}(E_N-m_\Delta)C^A_4
+m_N C^A_5-\frac{C^A_6}{m_N}p^j\left(\sum_{k=1}^3 p^k\right)\Bigg] ,
\label{S1} \\
S_1^A(\mathbf{q};4)& =& \sum^3_{\sigma=1}\Pi^A_\sigma({\bf 0},-\mathbf{q};\Gamma_4;4) 
= B\nonumber \\
&& \sum_{k=1}^3 p^k\Bigg[C^A_3+\frac{m_\Delta}{m_N}C^A_4
+\frac{E_N-m_\Delta}{m_N}C^A_6\Bigg], \\
\label{S1_4}
S_2^A(\mathbf{q};j)&=&\sum^3_{\sigma\ne k=1}\Pi^A_\sigma({\bf 0},-\mathbf{q};\Gamma_k;j)= 
i \frac{3 A}{2}\Bigg[\left(\sum_{k=1}^3 p^k\right) \nonumber \\
&&\hspace{-1.5cm}
\left(\delta_{j,1}(p^2-p^3)+
\delta_{j,2}(p^3-p^1)+\delta_{j,3}(p^1-p^2)\right)C^A_3\Bigg], \\
\label{S2}
S_3^A(\mathbf{q};j)&=& 
\Pi^A_3({\bf 0},-\mathbf{q};\Gamma_3;\mu) 
-\frac{1}{2}\biggl[\Pi^A_1({\bf 0},-\mathbf{q};\Gamma_1;\mu)
\biggr .\nonumber \\
&&\hspace*{-1.9cm}\biggl .+\Pi^A_2({\bf 0},-\mathbf{q};\Gamma_2;\mu)\biggr]
= i A\Bigg[ \frac{9}{4}\left(
\delta_{j,1}p^2p^3-\delta_{j,2}p^1p^3\right)C^A_3\Bigg],
\label{S3}
\eeq
\normalsize
where  $j=1,2,3$ and
\beq
A&=&\frac{B}{(E_N+m_N)} \quad,\nonumber \\
B&=&\sqrt{\frac{2}{3}}\frac{\sqrt{\left(E_N+m_N\right)/E_N}}{3m_N} \quad.\eeq

As can be seen, $S_2^A({\bf q};j)$ and $S_3^A({\bf q};j)$ isolate the
suppressed form factor $C_3^A$ for different combinations of
lattice momentum vectors as compared to $S_1^A$ that also involves $C_3^A$.
 Since here we are only interested in the dominant
form factors $C_5^A$ and $C_6^A$, we use only $S_1^A({\bf q};\mu)$.
We denote the ratio constructed  analogously  to $R^A_\sigma$, but with
the optimal sink, by $R_{N\Delta}^A$.
The linear combination $S_1^A({\bf q};\mu)$ turns out to  be
 the suitable one also for the calculation of the
from factor $G_{\pi N \Delta}$, defined in Eq.~(\ref{g_piND}).
 Again replacing the three-point function 
 $\langle G^{\Delta A^3_\mu N}_{\sigma} (t_2, t_1 ;
{\bf p}^{\;\prime}, {\bf p}; \Gamma) \rangle$ with the corresponding
pseudoscalar three-point function
 $\langle G^{\Delta P^3 N}_{\sigma} (t_2, t_1 ;
{\bf p}^{\;\prime}, {\bf p}; \Gamma) \rangle$
in Eq.~(\ref{RND-ratio}), we obtain the ratio $R^P_{\sigma}$,
which at large Euclidean times becomes
time-independent. Fitting in the plateau region yields
  $\Pi_{\sigma}^P ({\bf 0},-{\bf q}\; ;
\Gamma ;\gamma_5)$, 
related to the $\pi N\Delta$ form factor via the relation
\beq  
\Pi_\sigma^P ({\bf 0}, -{\bf q}\; ;
\Gamma_4 ;\gamma_5) &=& \\ \nonumber  
&\>&\hspace*{-3cm}\sqrt{\frac{2}{3}}\sqrt{\frac{E_N+m_N}{E_N}} \; 
\frac{q_\sigma}{6 m_N} \frac{f_\pi m_\pi^2} 
{2 m_q (m_\pi^2 + Q^2)} \; G_{\pi N \Delta} (Q^2)  
\eeq 
The optimal combination gives
\beq  
 S^P_{N\Delta}({\bf q}\; ;\; \gamma_5)&=& \sum_{\sigma=1}^3
\Pi_\sigma^P ({\bf 0},-{\bf q}\; ;
\Gamma_4 ;\gamma_5) = \sqrt{\frac{2}{3}}\sqrt{\frac{E_N+m_N}{E_N}} \nonumber\\   
& \;&\hspace*{-1.5cm} 
\left[\frac{q_1 + q_2 + q_3}{6 m_N} \frac{f_\pi m_\pi^2} 
{2 m_q (m_\pi^2 + Q^2)} \right]\; G_{\pi N \Delta} (Q^2) ,
\label{PND optimal}
\eeq 
from which $G_{\pi N\Delta}$ can be determined.

In order to evaluate $G_{\pi NN}$ and $G_{\pi N\Delta}$, we need to know $m_q$.
The
renormalized quark mass can be defined  via
the AWI given in Eq.~(\ref{quark mass}). 
On the lattice the AWI has corrections, which
in the case of Wilson fermions are of order $a$. For domain
wall fermions (DWF), the 
divergence of the four-dimensional vector axial current has an additional
term that goes to zero as the fifth dimension goes to infinity~\cite{shamir}.
For non-singlet matrix elements at low energies,  this additional term  
shifts the quark mass by an additive constant known as 
the  residual quark mass~\cite{Blum}. Such a simple shift is valid
 up to order $a^2$. Provided that the fifth dimension is large enough, the
 residual mass maybe considered a small correction. The
size of the fifth dimension was adjusted 
by  requiring the residual mass to be small compared to the pion mass.
 The criterion used to fix the fifth dimension to 16
is that the residual mass is  smaller than 10\% of the
quark mass~\cite{renner}.  We adopt the same criterion
in this work and take the  fifth dimension to be 16.
Therefore we assume that corrections due to the residual mass are small
and we  calculate the renormalized
quark mass  using the AWI 
taking matrix elements between a pion zero momentum state
and the vacuum. 
The initial state
with the pion quantum numbers is created 
using the axial-vector current
$\tilde{A}^3_4$  as an interpolating field.  The quantities with the tilde  denote
operators  that 
are built from smeared quark
fields obtained from point quark fields $\psi(x)$
as described in the next subsection.  The pion-vacuum matrix element
of the axial-vector current is given by the two-point
function
 \beq
C^{A}_{LS}(t) 
 = \sum_{{\bf x}}  \; \langle \Omega
|\;T\;\left( A_4^3 ({\bf x},t)
 \tilde{A}^3_4 ({\bf 0},0)\right) \; |  \Omega\;\rangle 
\eeq
and the pion-vacuum matrix element of the pseudoscalar density is given by 
\beq
C^{P}_{LS}(t) 
 = \sum_{{\bf x}}  \; \langle \Omega
|\;T\;\left( P^3 ({\bf x},t)
 \tilde{P}^3 ({\bf 0},0)\right) \; |  \Omega\;\rangle \quad.
\eeq
The subscript $L$ denotes that the axial-vector current 
and pseudoscalar density
are constructed using local quark fields unlike the interpolating fields
$\tilde{A}^3_4$ and $\tilde{P}^3$ that use smeared quark fields .
 To cancel the overlaps of our initial pion state with the vacuum 
we form the ratio
\beq
m_{\rm eff}^{\rm AWI}(t) =\frac{m_\pi}{2}\frac{Z_A}{Z_P}
\frac{C^{A}_{LS} (t)}{C^{P}_{LS}(t)}
\sqrt{\frac{C^{P}_{SS} (t)}{C^{A}_{SS}(t)}}
\label{meff}
\eeq
using, in addition to local-smeared (LS) two-point functions, 
the smeared-smeared 
two-point functions $C^{A}_{SS}$ and $C^{P}_{SS}$.
We look for a plateau in the
large Euclidean time behavior of the effective mass $m_{\rm eff}(t)$, 
which determines $m_q$.
The factors $Z_A$ and $Z_P$ are
the renormalization constants for the local axial-vector and pseudoscalar
currents, respectively. 
We note that $Z_P$ is only needed for the determination
of the renormalized quark mass. This dependence cancels in 
all physical quantities presented in this work,
which are therefore  independent of the value we use for $Z_P$.
The evaluation of $m_q$
 together with the determination of $f_\pi$ allows us to evaluate
$G_{\pi NN}$ and $G_{\pi N\Delta}$. 
The pion decay constant $f_\pi$ 
is determined from the large Euclidean 
time behavior of the ratio
\beq
f_\pi^{\rm eff}(t) = Z_A \sqrt{\frac{2}{m_\pi}}
\frac{C^{A}_{LS} (t)}{\sqrt{C^{A}_{SS}(t)}}
\; e^{m_\pi t/2} \quad.
\label{fpi}
\eeq
For large $t$, the above quantity becomes $t$ independent and the plateau
value gives $f_{\pi}$.

For the evaluation of all N to $\Delta$ matrix elements in the
case of  Wilson fermions, we use the 
sequential propagators already computed in 
our study of  the electromagnetic transition
 form factors~\cite{PRL_quenched,ND dynamical}.
However 
in the hybrid scheme, additional propagators are calculated to improve
the statistical errors beyond those of  our previous 
work~\cite{ND dynamical, PRL_axial} and to check for finite volume effects.
We summarize in Tables~\ref{Table:params Wilson} and \ref{Table:params hybrid}
the details of the calculation.   
All the hadron masses given in Table~\ref{Table:params hybrid} are computed
using domain wall valence quarks and MILC configurations for the sea quarks.
The value of the valence domain wall quark mass, $m_q^{DW}$,
 was determined by tuning the pion
mass calculated with domain wall fermions to be the same as
the lowest
mass pion in the staggered formulation~\cite{renner}.

For Wilson fermions,
we convert dimensionless lattice quantities to physical units by setting
the lattice spacing using the nucleon mass at the physical limit. 
The value of $a$ extracted from the nucleon mass
is given in Table~\ref{Table:params Wilson}
and it
is consistent with the value extracted  using the Sommer scale $r_0$.
The dynamical Wilson configurations at $\kappa=0.1575$ and $0.1580$ 
were generated by the $T\chi L$ 
collaboration~\cite{TchiL} and at $\kappa=0.15825$ by the DESY-Zeuthen
group~\cite{Carsten}.
For the hybrid calculation we use the scale 
  extracted from
heavy meson spectroscopy as determined by the MILC collaboration~\cite{MILC_a}.
 As can be seen in Table~\ref{Table:params hybrid} in the hybrid approach
we consider lattices with  temporal extent 
 $32$ and $64$. Temporal extent $32$ is obtained by
using Dirichlet boundary conditions (b.c.) in the temporal direction
to cut into half the original MILC lattices
when we calculate the domain wall quark propagator. This was the procedure
adopted in our previous evaluation of N to $\Delta$ axial form 
factors~\cite{ND dynamical} due to the limited computer resources.
 In this work we  present, in addition, results
using the full temporal extent of the MILC lattices 
with antiperiodic b.c. in the temporal direction
 consistent with
what is used in the
simulation of the  configurations.
Antiperiodic b.c. in the temporal direction are also used in the case of Wilson
fermions.

\begin{widetext}

\begin{table}
\begin{center}
\begin{tabular}{ccccc}
\hline \multicolumn{5}{c}
{Wilson fermions}\\
\hline No. of confs & $\kappa$ & $m_\pi$~(GeV) & $m_N$~(GeV) & $m_\Delta$ (GeV)\\ \hline
\multicolumn{5}{c}{Quenched $32^3\times64$, $\beta=6.0,~~a^{-1}=2.14(6)$~GeV} \\
\hline 200 &0.1554 &0.563(4) &1.267(11) &1.470(15)\\
 200 &0.1558 &0.490(4)&1.190(13) & 1.425(16)\\
  200& 0.1562 &0.411(4) &1.109(13) &1.382(19)\\
  &$\kappa_c$ =0.1571& 0.& 0.938(9) &\\
  \hline
  \multicolumn{5}{c}{Unquenched\cite{TchiL} $24^3\times40$,$\beta=5.6,~~a^{-1} = 2.56(10)$~GeV}\\
  \hline
  185 &0.1575 &0.691(8) &1.485(18) & 1.687(15)\\
  157 &0.1580 &0.509(8) &1.280(26) & 1.559(19)\\
  \hline
  \multicolumn{5}{c}{Unquenched\cite{Carsten} $24^3\times32$,$\beta=5.6,~~a^{-1} = 2.56(10)$~GeV}\\
  \hline
  200 & 0.15825 &0.384(8) &1.083(18) & 1.395(18)\\
  &$\kappa_c$ = 0.1585 &0. &0.938(33) &\\
  \hline
\end{tabular}
\end{center}
\caption{Parameters for the calculations using  Wilson fermions}
\label{Table:params Wilson}
\end{table}
\begin{table}
\begin{center}
\begin{tabular}{cccccccc}
\hline \multicolumn{8}{c}{
Hybrid action~~~$a^{-1}=1.58$~GeV~\cite{MILC_a}}\\
\hline No. of confs&Volume & $(am_{u,d})^{\textrm{sea}}$ &
$(am_{s})^{\textrm{sea}}$ & $(am_q)^{DW}$ & $m_{\pi}$ (GeV) &
$m_N$ (GeV) & $m_\Delta$ (GeV)\\
\hline
150&$20^3\times 32$ & 0.03 &0.05 &0.0478 & 0.606(2) & 1.392(9)  & 1.670(22)\\
150&$20^3\times 32$ & 0.02 &0.05 &0.0313 & 0.502(4) & 1.255(19) & 1.567(25)\\
118&$28^3\times 32$ & 0.01 &0.05 &0.0138 & 0.364(1) & 1.196(25) & 1.561(41)\\
200&$20^3\times 64$ & 0.03 &0.05 &0.0478 & 0.594(1) & 1.416(20) & 1.683(22)\\
198&$20^3\times 64$ & 0.02 &0.05 &0.0313 & 0.498(3) & 1.261(17) & 1.589(35)\\
100&$20^3\times 64$ & 0.01 &0.05 &0.0138 & 0.362(5) & 1.139(25) & 1.488(71)\\
150&$28^3\times 64$ & 0.01 &0.05 &0.0138 & 0.357(2) & 1.210(24) & 1.514(41)\\
\hline
\end{tabular}
\end{center}
\caption{Parameters for the calculationz using the hybrid action}
\label{Table:params hybrid}
\end{table}
\end{widetext}

\section{Extraction of observables}
\subsection{Ground state dominance and noise reduction}
As we already pointed out, in order to extract physical matrix elements, 
we must first  evolve in Euclidean time
to create the hadronic state of interest. In this work, the
hadronic states 
of interest  are the pion, the nucleon and the $\Delta$ states.
To create the initial states with  the pion quantum
numbers,  we use the temporal component of the
 axial-vector current, and  for the nucleon and $\Delta$,
the interpolating fields given in Eqs.~(\ref{nucleon field}) 
and (\ref{Delta field}) 
respectively.  The length of the time evolution
required to obtain the true pion, nucleon and $\Delta$ eigenstates
depends on our choice of the initial state. It is well known that
if one constructs a hadron initial state using smeared quark fields
 instead of localized ones, the convergence to the hadron eigenstate is
very much improved. Therefore, in this work, we always 
smear the quark fields in a gauge invariant way using the so called
Wuppertal or Gaussian smearing~\cite{Wuppertal}. In this scheme
the smeared quark field, $\tilde{\psi}({\bf x},t)$, is obtained
from the localized field, $\psi({\bf z},t)$, via
\be
\tilde{\psi}({\bf x},t) = \sum_{\bf z} F({\bf x},{\bf z};U(t)) \psi({\bf z},t)
\quad.
\ee
The gauge invariant smearing function is constructed from the
hopping matrix $H$:
\be
F({\bf x},{\bf z};U(t)) = (1+\alpha H)^n({\bf x},{\bf z};U(t)),
\ee
where 
\be
H({\bf x},{\bf z};U(t))= \sum_{i=1}^3 \biggl( U_i({\bf x},t)\delta_{{\bf x,y}-i} +  U_i^\dagger({\bf x}-i,t)\delta_{{\bf x,y}+i}\biggr).
\ee
The parameters 
for the Wuppertal smearing 
are determined by requiring that
 the nucleon state dominates the two-point correlator
 for the shortest time evolution.
We find that $\alpha=4$ and $n=50$ are  optimal parameters. 
Although smearing improves ground state dominance,
it introduces gauge noise increasing the
errors on the extracted effective masses in particular when 
applied  both to the source and 
to the sink. An efficient way to reduce the
 ultraviolet fluctuations is to smooth the gauge fields 
that enter the hopping matrix $H({\bf x},{\bf z};U(t))$.
It was shown in Ref.~\cite{NN lattice} that 
hypercubic smearing~\cite{HYP} on these gauge links
reduces gauge noise and tends to also improve ground state dominance.
In the case of domain wall fermions, HYP smearing is in fact  applied
to all the gauge links so as to accelerate
the convergence of the bi-conjugate gradient method used to evaluate
the inverse of the fermionic matrix.
In the quenched case, HYP smearing is not used because  self averaging 
is more effective on larger lattices.  In the case of
 dynamical Wilson fermions,
the simulations are done on smaller lattices
causing gauge noise, and HYP smearing needs to be applied to the gauge fields 
that enter the hopping matrix $H({\bf x},{\bf z};U(t))$.

\subsection{Plateaus and Overconstrained Analysis}
In this subsection we describe  the analysis of the lattice measurements
that lead to the  extraction of physical quantities.

\begin{figure}[h]
\epsfxsize=8truecm
\epsfysize=10truecm
\mbox{\epsfbox{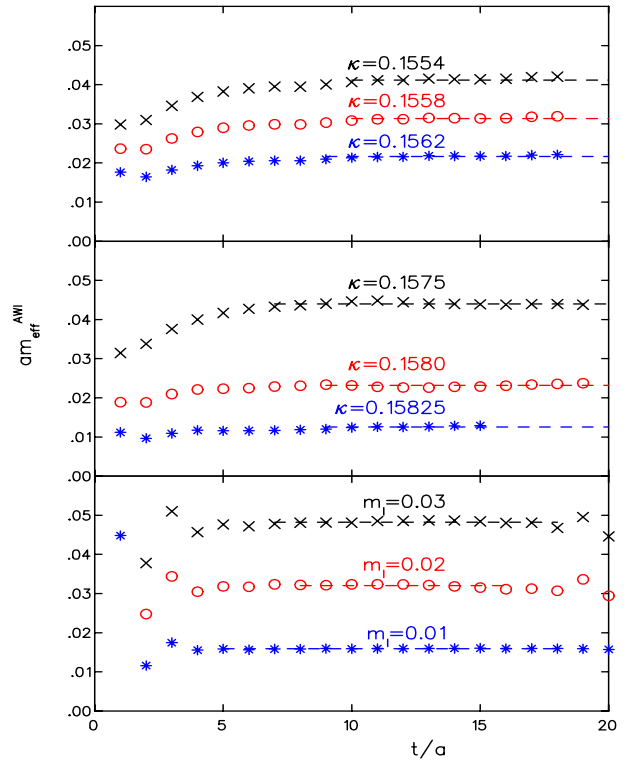}}
\caption{The effective quark mass $m_{\rm eff}^{\rm AWI}(t)$
  defined in Eq.~(\ref{meff}) as a function of  time, both in lattice units. 
The upper graph is for the quenched case, the middle graph for
dynamical Wilson fermions  and the lower graph for the hybrid scheme.
The dashed lines span the range of fitted points and show the extracted 
value of $m_q$ in lattice units.
}
\label{fig:meff}
\end{figure}

\begin{figure}[h]
\epsfxsize=8truecm
\epsfysize=8truecm
\mbox{\epsfbox{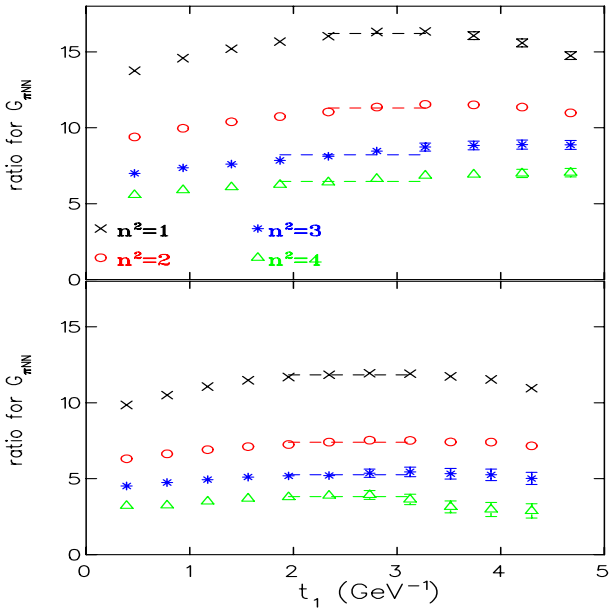}}
\caption{The ratio $R^P$ used to extract $G_{\pi NN}$  for the
four lowest values of $Q^2$. The upper graph is for the quenched theory
at $\kappa=0.1558$  ($m_\pi=0.49$~GeV) and the lower graph
for dynamical Wilson fermions at $\kappa=0.1575$ ($m_\pi=0.69$~GeV).
The dash lines are fits
to the plateaus and span the range of fitted points.}
\label{fig:ratios GPNN}
\end{figure}

\begin{figure}[h]
\epsfxsize=8truecm
\epsfysize=10truecm
\mbox{\epsfbox{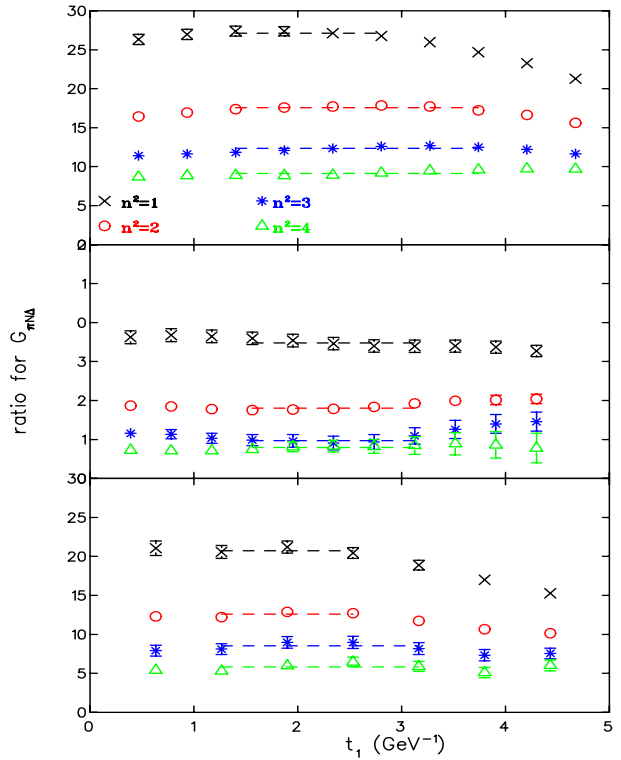}}
\caption{The ratio $R_{N\Delta}^P$ used to extract  $G_{\pi N\Delta}$  for the
four lowest values of $Q^2$. The upper graph is 
for the quenched theory, the middle graph
for dynamical Wilson fermions and the lower graph for hybrid scheme
 for a pion of mass about 500 MeV (intermediate value).
The dash lines are fits
to the plateaus and span the range of fitted points.
}
\label{fig:ratios GPND}
\end{figure}
\begin{figure}[h]
\epsfxsize=8truecm
\epsfysize=8truecm
\mbox{\epsfbox{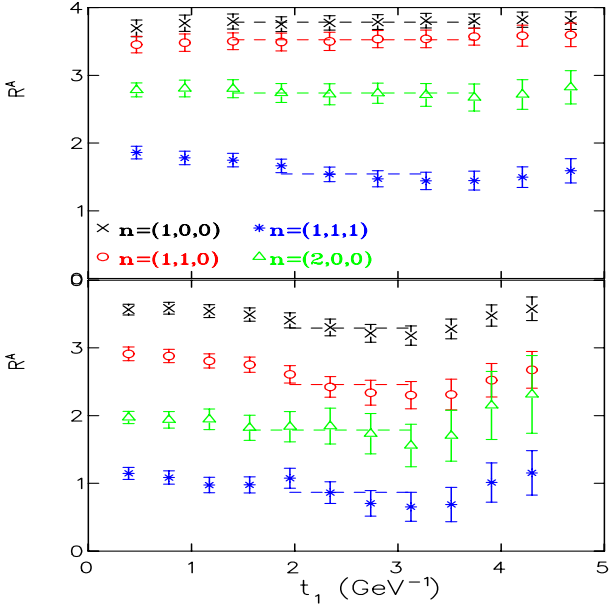}}
\caption{The ratio $R^A$ defined in Eq.~(\ref{R-ratio}) with the optimal
nucleon sink $S^A$
from which $G_A$ and $G_p$ are extracted
for the four lowest 
momentum vectors  ${\bf q}=(1,0,0)2\pi/L_s$, ${\bf q}=(1,1,0)2\pi/L_s$,
${\bf q}=(1,1,1)2\pi/L_s$ and ${\bf q}=(2,0,0)2\pi/L_s$, where $L_s$ is
the spatial size of the lattice.
The upper graph is for the quenched theory at intermediate pion mass 
($\kappa=0.1558$) and the lower for two dynamical Wilson quarks at the
heaviest mass ($\kappa=0.1575$).
The dash lines are fits
to the plateaus and span the range of fitted points.
}
\label{fig:ratios GA HA}
\end{figure}

\begin{figure}[h]
\epsfxsize=8truecm
\epsfysize=8truecm
\mbox{\epsfbox{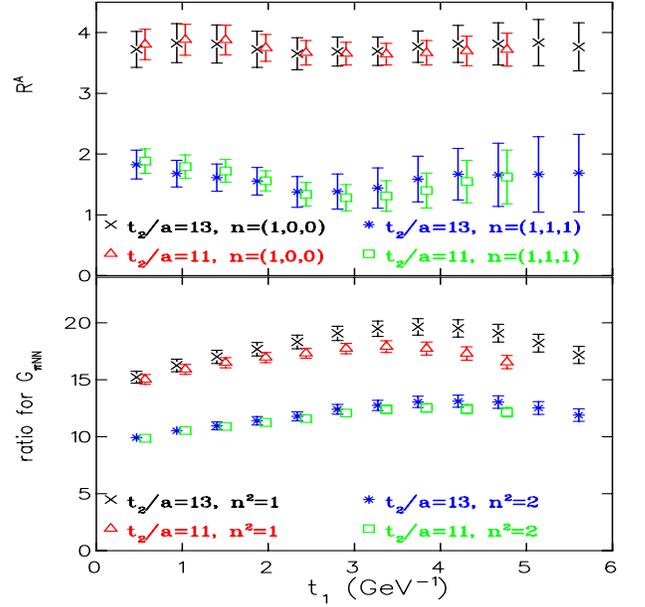}}
\caption{The upper graph shows $R^A$, from which $G_A$ and $G_p$ are extracted,
for momentum vectors ${\bf q}=(1,0,0)2\pi/L_s$ and ${\bf q}=(1,1,1)2\pi/L_s$.
The lower graph shows  the ratio $R^P$, from which  $G_{\pi NN}$ is determined, for momentum transfer squared
${\bf q}^2=(2\pi/L_s)^2$ and ${\bf q}^2=2(2\pi/L_s)^2$.
Results on these quantities are shown as a function of $t_1$
for the quenched theory  for sink-source separations, $t_2/a=13$ (crosses),
and $t_2/a=11$ (open triangles) at the
smallest quark mass ($\kappa=0.1562$).  }
\label{fig:plateaus compare}
\end{figure}

\begin{figure}[h]
\epsfxsize=8truecm \epsfysize=10truecm 
\mbox{\epsfbox{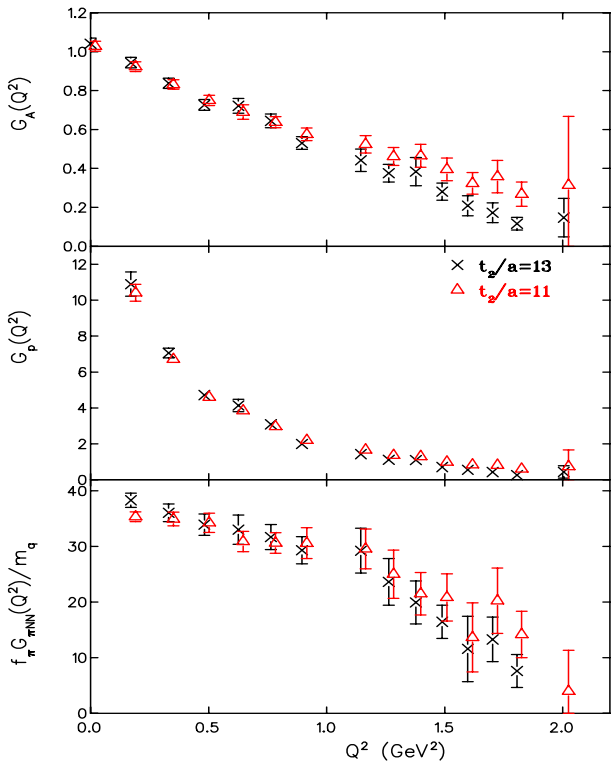}}
\caption{The upper graph shows $G^A(Q^2)$, the middle graph    $G_p(Q^2)$ 
and the lower graph $f_\pi G_{\pi NN}/m_q$ as a function of $Q^2$
in the quenched theory for sink-source separations, $t_2/a=13$ (crosses),
and $t_2/a=11$  at the
smallest quark mass ($\kappa=0.1562$). }
\label{fig:quenched compare}
\end{figure}
\begin{figure}[h]
\epsfxsize=8truecm
\epsfysize=10truecm
\mbox{\epsfbox{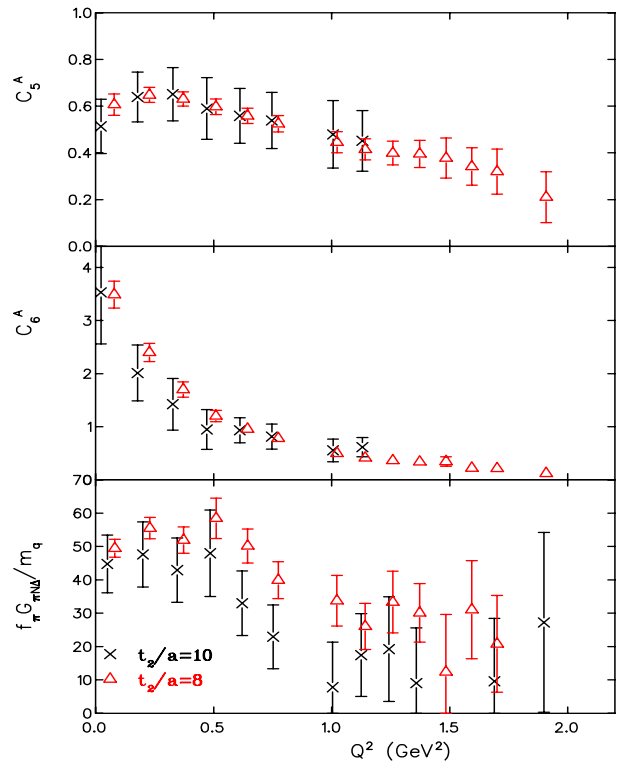}}
\caption{The upper graph shows $C_5^A(Q^2)$, 
the middle graph $C_6^A(Q^2)$ and the lower graph
 $f_\pi G_{\pi N\Delta}/m_q$ as a function of $Q^2$
in the hybrid scheme for sink-source separations, $t_2/a=10$ (crosses),
and $t_2/a=8$ at the
smallest quark mass, namely $m_l=0.01$. }
\label{fig:MILC compare}
\end{figure}

\begin{figure}[h]
\epsfxsize=8truecm
\epsfysize=10truecm
\mbox{\epsfbox{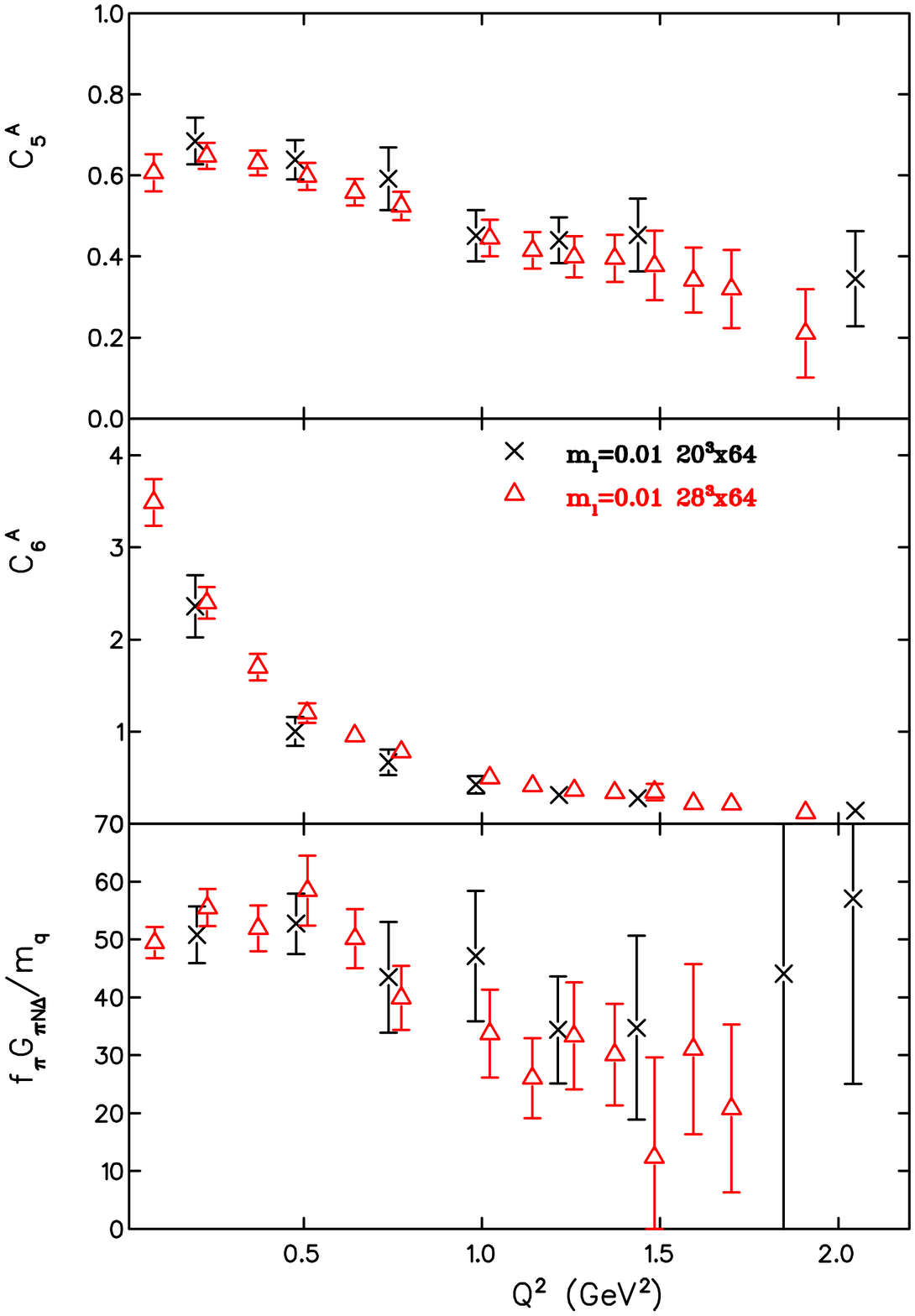}}
\caption{The upper graph shows $C_5^A$, the middle graph $C_6^A$  and the lower
graph
$G_{\pi N\Delta}$  as a function of $Q^2$
 in the hybrid approach for spatial volumes, $20^3$ (crosses) 
and $28^3$ (open triangles) at the
smallest quark mass, $m_l=0.01$. }
\label{fig:MILC volume}
\end{figure}

The mass of the lowest
hadron state for a given set of quantum numbers is
the simplest quantity to calculate on the lattice,  since it  requires only
the computation of two-point
functions. In this work we need, besides the pion, the nucleon and 
the $\Delta$ mass,
which are straightforward to determine,
the renormalized quark mass. This is 
evaluated by taking matrix elements of the AWI as discussed in Section III.
The effective mass $m_{\rm eff}^{\rm AWI}(t)$ defined in 
 Eq.~(\ref{meff}) 
becomes time-independent if  $t$ is large enough so that
the pion ground state dominates (plateau region).
We show in Fig.~\ref{fig:meff} $m_{\rm eff}^{\rm AWI}(t)$ 
as a function of time, both in lattice units. We consider
all  three values of
the bare quark mass for each of  the
three types of simulations that we
use in this work, namely  the  quenched approximation,
two  dynamical Wilson fermions and  the hybrid scheme.
As can be seen, in all cases, allowing for an initial time evolution, the
effective mass becomes time independent. Fitting to a constant
in the plateau region determines
$m_q$. In the case of DWF, the
extracted value  is expected to be  the same
  as $m_q^{DW}$ used in the domain wall
Dirac matrix and given in 
Table~\ref{Table:params hybrid}. Any differences 
are attributed  to an additive residual mass
that provides a measure of the chiral symmetry breaking due to the
finite extent of the fifth dimension. 
In the case of Wilson fermions it is known 
that the  AWI has corrections of ${\cal O}(a)$. The value
of $m_q$ therefore carries systematic errors, which can be large 
as we approach the chiral limit, since corrections that
appear in the right hand side of Eq.~(\ref{ward}) will dominate as the first
term decreases.  The results on $m_q$ will be discussed in the next
section.

For the evaluation of form factors we look for time independence
of the ratios $R^A$, $R^P$, $R_{N\Delta}^A$ 
and $R_{N\Delta}^P$ constructed
from three-point functions and appropriate combinations of 
two-point functions. We start with the ratio $R^P$, which in the large
 Euclidean limit goes to $S^P$
given in Eq.~(\ref{P optimal}),
from which  the $\pi NN$ form factor is determined. We show in
Fig.~\ref{fig:ratios GPNN} typical examples of the ratio
$R^P$  divided by $C(q_1+q_2+q_3)/2m_N$ and averaged
over all momentum directions that lead
to the same $Q^2$ value.
Such an averaging is also done
 in our overconstrained analysis 
described below.
The ratio is shown  as a function of $t_1$ in physical units for
the four lowest $Q^2$-values
 both in the quenched theory at the intermediate
quark mass and for dynamical Wilson fermions at the heaviest quark mass.
 When the time separation from the source and sink is large enough 
so that the nucleon state dominates,  this averaged ratio
is time independent.
  When this happens the quantity plotted  
in Fig.~\ref{fig:ratios GPNN}
corresponds to  $f_\pi m_\pi^2 G_{\pi NN}/[2 m_q (Q^2+m_\pi^2)Z_P ]$.
Note that,
since to obtain  the renormalized mass we divide by $Z_P$,  the
$Z_P$ factors cancel and therefore we do not need to
know $Z_P$. As already pointed out, this is true for all physical
quantities that we calculate in this work. The dashed lines show
both the range used for the fit and the value of the plateau.
In Fig.~ \ref{fig:ratios GPND} we show the corresponding average of
 the ratio $R_{N\Delta}^P$ that, in the large time limit, leads to $S^P_{N\Delta}$
defined in Eq.~(\ref{PND optimal}) and to the determination
of   the $\pi N\Delta$ form factor.
Here we show results for pions of mass of about 500~MeV (intermediate value)
in each of the three types of simulations considered in this work.
As can be seen the quality of the plateaus in all cases is good
enough to allow us to fit to
a constant within the
plateau range  with a $\chi^2$/degrees of freedom $(d.o.f)\stackrel{<}{\sim} 1$.
This leads to a good determination of
 the $\pi NN$ and $\pi N\Delta$ form factors.

In Fig.~\ref{fig:ratios GA HA} we show the ratio $R^A$ defined in 
Eq.~\ref{R-ratio} for
the optimal combination that, for large $t_1$ and $t_2-t_1$ time intervals,
leads to Eq.~(\ref{SA optimal}). We show results
 for the four lowest values of the lattice
momentum vector, ${\bf q},$ since,
in this case, the ratio depends on the 
momentum vector and not just its magnitude. 
We note that this is not what is actually fitted, 
since in our overconstrained analysis we consider all lattice vectors ${\bf q}$
that result in the same $Q^2$ value. However Fig.~\ref{fig:ratios GA HA}
gives an idea of the quality of the plateaus that
are used in the overconstrained analysis
to extract the nucleon axial form factors.
Similar plateaus are obtained for the ratio $R^A_{N\Delta}$ needed to extract
 the N to $\Delta$ axial form factors. 
Again the quality of the data allows identification
of a plateau region to which to perform a fit to a constant
to extract the matrix element we are interested in.

The overconstrained analysis uses all the stochastically independent
lattice measurements that contribute at a given $Q^2$ when extracting the
form factors~\cite{svdjohn}. 
This is done by solving the overcomplete set of equations
\be
P({\bf q};\mu)= D({\bf q};\mu)\cdot F(Q^2)
\ee
where $P({\bf q};\mu)$ are lattice measurements of appropriately
defined ratios. For concreteness, let us consider the analysis
for the nucleon axial
form factors. In this case $P({\bf q};\mu)$ is the ratio $R^A$
given in Eq.~(\ref{R-ratio}) having statistical errors
$w_k$. 
The vector $F$
contains  the form factors:
\be F(Q^2) =  \left(\begin{array}{c} G_{A}(Q^2) \\
                                  G_{p}(Q^2) 
                                   \end{array}\right) \quad.
\ee
If $N$ is the number of current
directions and momentum vectors contributing to
a given $Q^2$ then  $D$ is an $N\times 2$ matrix which depends on
kinematical factors. We extract the form factors by
minimizing
\be
\chi^2=\sum_{k=1}^{N} \Biggl(\frac{\sum_{j=1}^2 D_{kj}F_j-P_k}{w_k}\Biggr)^2
\ee
using the singular value decomposition of $D$.
Therefore we do not actually fit to the plateaus shown in 
Fig.~\ref{fig:ratios GA HA} for each momentum vector but combine
all momentum vectors in the overconstrained analysis.
A similar analysis is done for the determination of all the other
form factors.

\subsection{Fixing the source-sink time separation}
All the results shown in Figs.~\ref{fig:ratios GPNN}, \ref{fig:ratios GPND}
and \ref{fig:ratios GA HA} are obtained keeping
the source-sink separation, $t_2$,  fixed.
In the quenched case we take $t_2/a=11$, 
for dynamical Wilson fermions we take $t_2/a=12$
and for the hybrid scheme we take $t_2/a=8$
so as to  keep the physical time separation approximately
constant at about 5~GeV$^{-1}$ or 1~fm. 
In order to ensure that this time separation is
large enough to isolate the nucleon and $\Delta$ states we must increase
the sink-source time separation and check
that the results remain unchanged. This check is carried out in the quenched
theory at the lowest quark mass and in the hybrid
scheme.
In both, 
we increase the source-sink separation by two time slices.

We choose to do this check for quenched  rather
than dynamical Wilson fermions since the errors are smaller
and we can therefore identify deviations more easily.
 We choose  the smallest
mass, since the smaller the mass the more severe is the contamination of
excited states. 
In Fig.~\ref{fig:plateaus compare} we show the ratio $R^A$ 
for the optimal nucleon source 
 $S^A$ of Eq.~(\ref{SA optimal}) 
for source-sink time separation $t_2/a=11$ and $t_2/a=13$.
Results are shown for the lowest momentum vector
${\bf q}=(1,0,0) 2\pi/L_s$ and for ${\bf q}=(1,1,1) 2\pi/L_s$,
where $L_s$ is the spatial extent of the lattice.
As can be seen the ratios yield consistent plateaus. In the same figure
we also show  the ratio  $R^P$ from which $G_{\pi NN}$ 
is extracted for the two lowest 
${\bf q}^2$-values. The only discrepancy arises at the lowest 
${\bf q}^2$ value, where the larger source-sink separation  
prodices a larger result.
At all higher
values of ${\bf q}^2$, 
the plateaus are however consistent as demonstrated for the second 
lowest value of ${\bf q}^2$. Given that the plateaus
for $R^A$ for both time separations
are consistent at all values of the momentum vectors, the
discrepancy seen in the case of  $R^P$ at the smallest ${\bf q}^2$ value may 
have a different origin.
We note  that $S^P$  is proportional to $q_1+q_2+q_3$.
As ${\bf q}\rightarrow 0$ extracting $G_{\pi NN}$  becomes ill-defined
and our statistical error in this case 
underestimates the true error. The effect of increasing $t_2$ on
the actual form factors can be seen in
 Fig.~\ref{fig:quenched compare}, where we show the nucleon axial form
factors $G_A(Q^2)$ and $G_p(Q^2)$ as well as $G_{\pi NN}(Q^2)$ 
extracted for sink-source
separations $t_2/a=11$ and $t_2/a=13$. 
As can be seen,
the results up to $Q^2\sim 1.5$~GeV$^2$
at the two time separations are within
error bars  with the only exception
the  value of $G_{\pi NN}$ at
the lowest $Q^2$ value,  which differs by about one standard deviation.
Differences by about one standard deviation in the results for
$G_A(Q^2)$ for $Q^2>1.5$~GeV$^2$ are, most likely,
 due to taking numerically the Fourier transform, which
 for large values of  $Q^2$,
becomes noisy, requiring more statistics.
Given this level of agreement at the smallest quark mass
we conclude
that, for the quenched case, a physical time distance of about 
5 GeV$^{-1}\sim 1$~fm is 
sufficient for ground state dominance and identification of
 a consistent  plateau region  with the hindsight that 
$G_{\pi NN}$  at the smallest $Q^2$ maybe underestimated
 by about one standard deviation. 

We next discuss the adequacy of the sink-source separation
 in  the hybrid approach. Pion
cloud contributions are expected to become important
for dynamical quarks as the quark mass decreases and one
must allow a large enough time separation for the pion cloud
to develop. Therefore it is important to ensure that the time
separation $t_2$ is large enough for dynamical quarks with the
smallest mass.
The results for the larger time separation 
are obtained using Dirichlet b.c. at the first time slice and at the midpoint
of the temporal direction cutting in half the lattice size whereas
for the smaller separation antiperiodic b.c. are used. 
We compare in Fig.~\ref{fig:MILC compare}
 the N to $\Delta$ form factors extracted for
 $t_2/a=10$ to those
obtained with sink-source time separation $t_2/a=8$. As can be seen
all the results
at the two time separations, including $G_{\pi N\Delta}$ at 
the lowest ${\bf q}^2$ value, are within
error bars. Given that we use the same number of configurations
for the two time  separations 
it is obvious that we have a big advantage for using the smaller separation
since errors are reduced by more
that a factor of two.
Given the level of agreement at the smallest quark mass 
for both quenched and hybrid results, combined with the advantage
of
smaller statistical errors, we conclude  that it suffices to  
 take $t_2\sim 5$~GeV$^{-1}$. Therefore all the results given
in the next section are obtained with this time separation. Furthermore 
results in the hybrid scheme  are obtained  using the
full temporal extent of the MILC lattices with
antiperiodic b.c. in the temporal direction.

\subsection{Volume dependence}
Another potential source of a systematic error 
is the spatial size of our lattices.
Given that for the quenched case we use a lattice of spatial size 
of about 3~fm 
we  expect finite volume effects to be negligible.
A rule of thumb is that finite volume effects are small if
 $L_sm_\pi \stackrel{\sim}{>}4-5$.
For all quark masses used in this work we have  $L_s m_\pi > 4.6$, except  
 for dynamical
Wilson fermions 
at the smallest quark mass where we have
$L_s m_\pi=3.6$.
Since we do not
have dynamical Wilson configurations on a larger volume we test for
finite size effects in the hybrid scheme for which,  at the smallest quark
mass, there are
 MILC configurations for $L_s=2.5$ and $L_s=3.5$ giving $L_sm_\pi=4.6$
and $L_s m_\pi=6.4$, respectively. 
In Fig.~\ref{fig:MILC volume}
 we show results for the N to $\Delta$ axial form factors 
$C_5^A(Q^2)$ and $C_6^A(Q^2)$ as well as
$G_{\pi N\Delta}(Q^2)$ 
for these two spatial
sizes. 
Results on the smaller lattice are consistent with results on the larger
lattice. This indeed shows that finite volume effects
are small for  $L_sm_\pi\stackrel{\sim}{>}4.5$.  Since for all our
quark masses, except the lightest mass dynamical 
Wilson fermions 
 $L_sm_\pi>4.6$, we expect finite volume effects to be small.
 We note, however,  that
a systematic study of volume effects that
would allow an extrapolation of our quantities to infinite volume
requires results using at least three different volumes. This  is beyond the scope of the present work.

Finally we  comment on the evaluation of the kinematical factors
in the expressions for the form factors, which involve the masses of 
the nucleon and $\Delta$ and their energies. The masses are evaluated
using two-point functions in the standard way. The energies are
calculated using the continuum dispersion relation,
$E=\sqrt{m^2+p^2}$, where $p^2= n(2\pi/L_s)^2,\> n=1,2,..$. 
One can compare results obtained using continuum dispersion relations
to those obtained with the  lattice dispersion relation
$\sinh^2(E)= \sinh ^2 m+\sum_{i=1,...,3}\sin^2 2\pi n_i/L_s$. 
We find that the mean value of the form factors is almost
unchanged. 
The $Q^2$-values are also very close for  $Q^2\stackrel{<}{\sim}1$~GeV$^2$.
 At larger momentum transfers
  the lattice dispersion relation shifts the $Q^2$ 
to smaller values.
 Using two-point functions to extract the energy
also yields consistent results for the form factors
albeit with larger errors. In what follows we will present
results as a function of $Q^2$ calculated using the continuum dispersion
relation.

\section{Results}

We first discuss results on quantities and  ratios for which
the renormalized quark mass is not required.
This eliminates one source of systematic error,
namely lattice artifacts on the value of
 $m_q$.
Furthermore, in general, ratios show weaker
dependence on quark mass.
For these reasons, they are more suited for comparison
with physical results.

\begin{figure}[h]
\epsfxsize=8truecm \epsfysize=5truecm
\mbox{\epsfbox{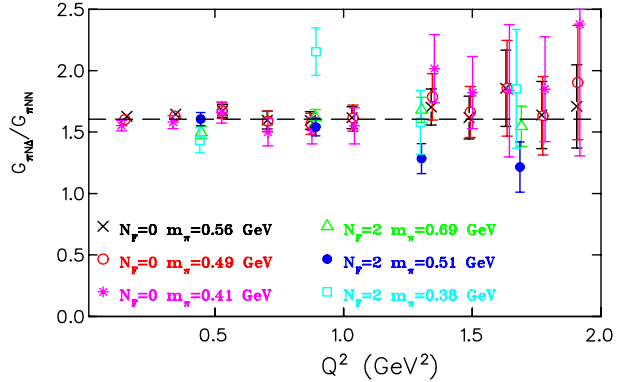}}
\caption{The ratio of form factors $G_{\pi N\Delta}(Q^2)/G_{\pi NN}(Q^2)$
 as a function of $Q^2$ for
Wilson fermions for the quenched theory, denoted by $N_F=0$, at 
$\kappa=0.1554 \> (m_\pi=0.56$ GeV), $\kappa=0.1558 \>(m_\pi=0.49$ GeV)
and $\kappa=0.1562 \>(m_\pi=0.41$ GeV) 
and for two dynamical Wilson quarks, denoted
by $N_F=2$, at $\kappa=0.1575 \>(m_\pi=0.69$ GeV)~\cite{TchiL},
 $\kappa=0.1580 \>(m_\pi=0.51$ GeV)~\cite{TchiL}
and $\kappa=0.15825\> (m_\pi=0.38$ GeV)~\cite{Carsten}. The dashed line
is the result of fitting  
 the quenched results to a constant, yielding  a value of 1.60(2).}
\label{fig:gpiNDovergpiNN}
\end{figure}

\begin{figure}[h]
\epsfxsize=8truecm \epsfysize=5truecm
\mbox{\epsfbox{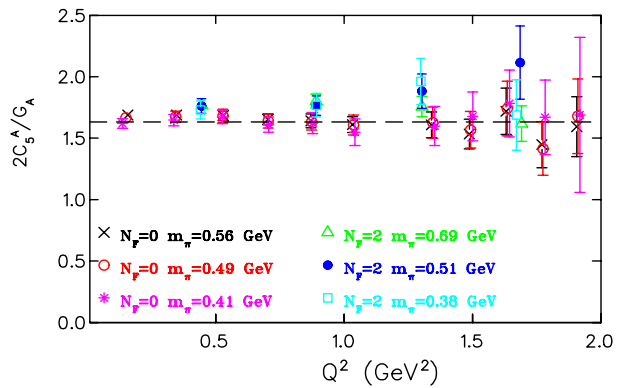}}
\caption{The ratio of $2C_5^A(Q^2)/G_A(Q^2)$ as a function of $Q^2$. 
The notation is the same as that of Fig.~\ref{fig:gpiNDovergpiNN}. 
Fitting   
 the quenched results to a constant yields  a value of 1.63(1).}
\label{fig:CA5overGA}
\end{figure}

\begin{figure}[h]
\epsfxsize=8truecm \epsfysize=5truecm
\mbox{\epsfbox{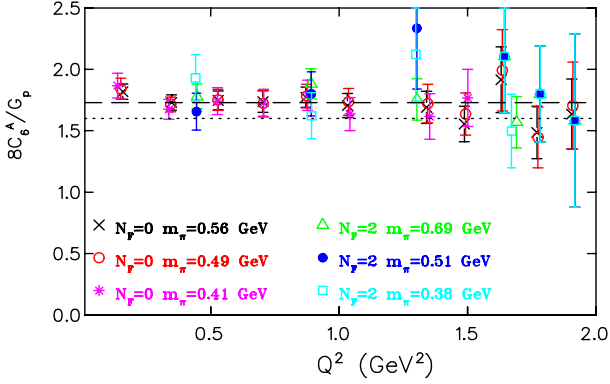}}
\caption{The ratio of $8C_6^A(Q^2)/G_p(Q^2)$ as a function of $Q^2$. 
The notation is the same as that of Fig.~\ref{fig:gpiNDovergpiNN}. 
Fitting   
 the quenched results to a constant yields  a value of 1.73(3).
The dotted line denotes the value of $1.60$ obtained by fitting
the ratio  $G_{\pi N\Delta}(Q^2)/G_{\pi NN}(Q^2)$.}
\label{fig:CA6overHA}
\end{figure}

\begin{figure}[h]
\epsfxsize=8truecm \epsfysize=5truecm
\mbox{\epsfbox{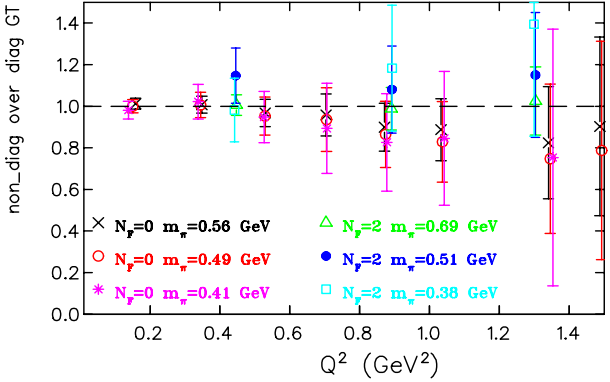}}
\caption{The ratio of Eq.~(\ref{GTR_ND}) to Eq.~(\ref{GTR}). The notation
is the same as that of Fig.~\ref{fig:gpiNDovergpiNN}.}
\label{fig:GTR_NDoverGTR}
\end{figure}

For the same lattice momentum vectors the $Q^2$ values
for the nucleon system differ from those in the $N-\Delta$
system.
In order to take ratios of form factors computed in these
two different systems
we interpolate the  form factors in the nucleon system
to the $Q^2$ value
of the $N-\Delta$ system. 
In Fig.~\ref{fig:gpiNDovergpiNN} we show the ratio of the form factors
$G_{\pi N\Delta}(Q^2)/G_{\pi NN}(Q^2)$ for quenched and two degenerate
flavors of dynamical Wilson quarks denoted by $N_F=0$ and $N_F=2$ 
respectively. 
As can be seen, this ratio is $Q^2$ independent and
shows no statistically significant quark mass dependence.
Fitting  the quenched results to a constant we obtain the value
of 1.60(2) shown by the dashed line. 
If we assume pole dominance for
the form factors $G_p(Q^2)$ and $C_6^A(Q^2)$ then the GTRs simplify to
the relations given in Eq.~(\ref{GTR simple}).
Taking the ratio of the diagonal and non-diagonal relations we find that 
 $G_{\pi N\Delta}(Q^2)/G_{\pi NN}(Q^2)=2C_5^A(Q^2)/G_A(Q^2)$. 
In Fig.~\ref{fig:CA5overGA}, we show the ratio $2C_5^A(Q^2)/G_A(Q^2)$,
 which is indeed
also $Q^2$ independent, and fitting to the quenched
data we find  the value of $1.63(1)$ shown by the dashed line.
Therefore,  on the level of ratios, the GTRs are
satisfied. We can use the relations given in Eq.~(\ref{GP&CA6})
for $G_p$ and $C_6^A$  to eliminate
$G_{\pi NN}$  and $G_{\pi N\Delta}$ 
in Eqs.~(\ref{GTR}) and (\ref{GTR_ND}) to obtain 
\beq
G_p(Q^2)=\frac{4m_N^2/m_\pi^2}{1+Q^2/m_\pi^2}\> G_A(Q^2) \nonumber \\
C_6^A(Q^2)=\frac{m_N^2/m_\pi^2}{1+Q^2/m_\pi^2}\> C_5^A(Q^2) \quad.
\label{pole dominance}
\eeq
These relations are again a manifestation of pion pole dominance.
 Taking  ratios, we find that  $8C_6^A(Q^2)/G_p(Q^2)$ should be
equal to the ratio  $2C_5^A(Q^2)/G_A(Q^2)$ and 
consequently to $G_{\pi N\Delta}(Q^2)/G_{\pi NN}(Q^2)$. 
As can be seen in Fig.~\ref{fig:CA6overHA},
we indeed find that also this ratio is constant as a function
of $Q^2$. Fitting  the quenched data to a constant we obtain 
 the value of  $1.73(3)$, shown by the dashed
line in the figure. This is  about 6\% larger than what
we find for the other two ratios. Therefore we conclude that
ratios based on the relations given in Eq.~(\ref{GTR simple}) 
are better satisfied
than ratios obtained using  Eq.~(\ref{pole dominance}).
Relaxing the assumption on pion pole dominance of $G_p$ and
$C_6^A$, we can consider directly the ratio of the non-diagonal to the diagonal
GTR given in Eqs.(\ref{GTR_ND}) and Eqs.(\ref{GTR}) respectively. As can
be seen in Fig.~\ref{fig:GTR_NDoverGTR}, the ratio is indeed consistent
with unity.

\begin{figure}[h]
\epsfxsize=8truecm
\epsfysize=5truecm
\mbox{\epsfbox{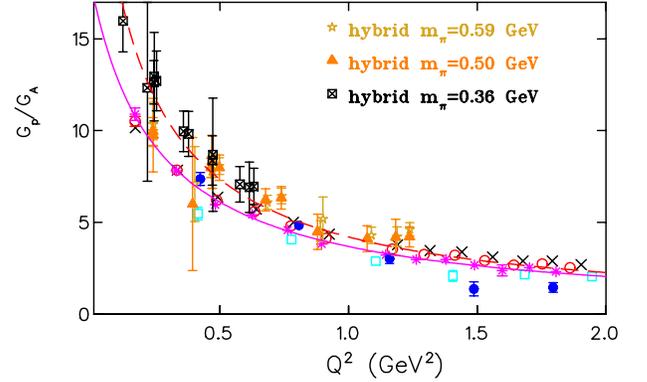}}
\caption{The ratio of nucleon axial form factors $G_p(Q^2)/G_A(Q^2)$ for
Wilson fermions for the quenched theory  
 and for two dynamical Wilson quarks using the same notation  as in 
Fig~\ref{fig:gpiNDovergpiNN}. 
We also show results from Ref.~\cite{LHPC07} obtained in the hybrid approach
using the same quark masses as the ones used
in this work, namely $m_l=0.03$~(star),
 $m_l=0.02$~(filled triangle) and $m_l=0.01$~(inscribed squares).
The dash line shows the expected behavior 
assuming pion pole dominance as given in Eq.~(\ref{pole dominance}), 
where for 
$m_\pi$ and $m_N$ we use the values
computed on the lattice at $\kappa=0.1562$. 
The solid curve 
is a fit to a monopole form of the quenched data at $\kappa=0.1562$. }
\label{fig:HAoverGA}
\end{figure}

\begin{figure}[h]
\epsfxsize=8truecm \epsfysize=5truecm
\mbox{\epsfbox{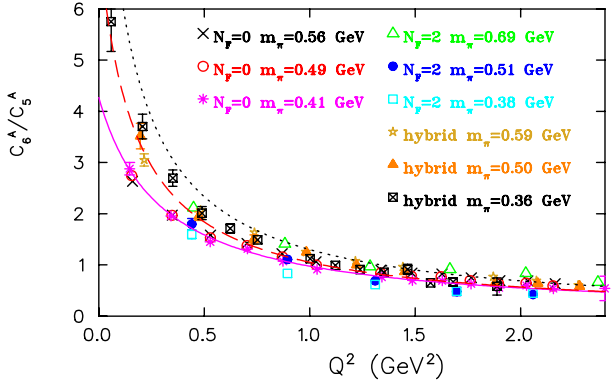}}
\caption{The ratio of $N$ to $\Delta$ axial transition 
 form factors $C_6^A(Q^2)/C_5^A(Q^2)$. 
The notation 
is the same as that of Fig.~\ref{fig:HAoverGA}.
 The dotted line  shows the prediction
of pion pole dominance predicted in Eq.~(\ref{pole dominance}) but for the 
hybrid case at the lightest quark mass.}
\label{fig:CA6overCA5}
\end{figure}

\begin{figure}[h]
\epsfxsize=8truecm \epsfysize=8truecm
\mbox{\epsfbox{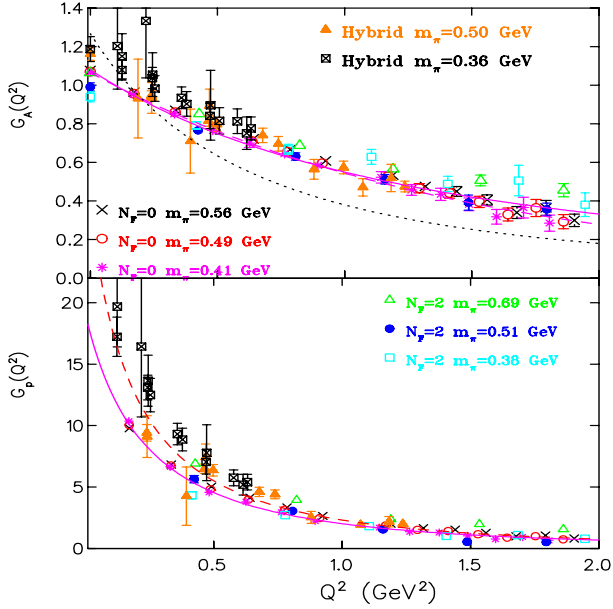}}
\caption{The upper graph shows $G_A(Q^2)$ and the 
lower graph $G_p(Q^2)$ as a function of $Q^2$. The solid curve
is a fit to a dipole form of the quenched results at $\kappa=0.1562$.The fit
to an exponential form  shown by the dashed line  falls on top. The 
dotted line shown in the upper graph
corresponds to a dipole form with axial mass 
$m_A=1.1$~GeV used to described experimental data. 
 The dashed line in th elower graph shows the result expected
from pion pole dominance in Eq.~(\ref{pole dominance}).
The solid line corresponds to Eq.~\ref{fit functions}.
 The rest of the notation is the same as that in
Fig.~\ref{fig:HAoverGA}.
Results in the hybrid
approach are from Ref.~\cite{LHPC07}.}
\label{fig:GAHA}
\end{figure}

\begin{figure}[h]
\epsfxsize=8truecm \epsfysize=8truecm
\mbox{\epsfbox{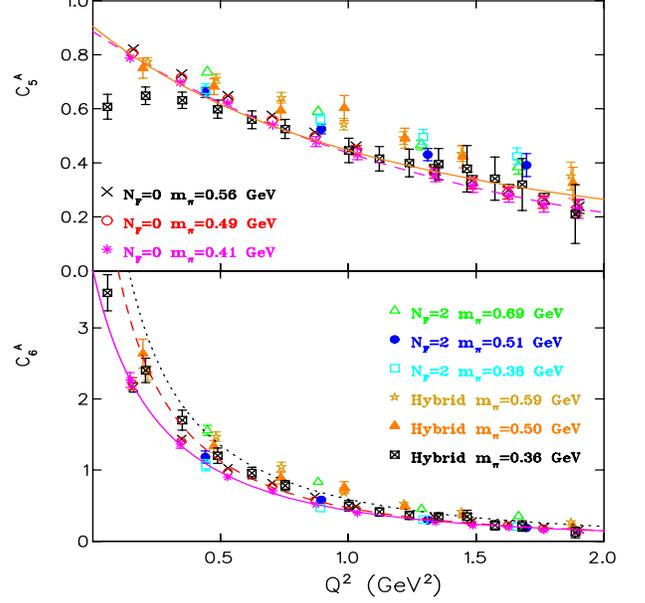}}
\caption{The upper graph shows $C_5^A(Q^2)$ and lower graph
 $C_6^A(Q^2)$ as a function of $Q^2$.
 The notation is the same as that in
Fig.~\ref{fig:HAoverGA}.}
\label{fig:CA5CA6}
\end{figure}

Having examined the $Q^2$-dependence
on the level of  ratios of GTRs in the nucleon and N-$\Delta$ 
systems we now discuss the $Q^2$ dependence of the form factors for
the two systems separately. 
In Fig.~\ref{fig:HAoverGA} we show the ratio of nucleon axial form factors
$G_p(Q^2)/G_A(Q^2)$ as a function of $Q^2$ for quenched and two degenerate
flavors of dynamical Wilson quarks. 
 Recent results from Ref.~\cite{LHPC07}
obtained in the hybrid scheme at the same quark masses as the ones
used in this work for the calculation of the $N$ to $\Delta$ form factors 
are also included.
In all cases, the ratio decreases with $Q^2$  confirming
the stronger $Q^2$-dependence  
expected for $G_p(Q^2)$ as compared to $G_A(Q^2)$.
A similar behavior is also observed for the corresponding ratio 
$C_6^A(Q^2)/C_5^A(Q^2)$ for N to $\Delta$ shown in
Fig.~\ref{fig:CA6overCA5}.
If pion pole dominance holds, then the ratios $G_p(Q^2)/G_A(Q^2)$ and 
$C_6^A(Q^2)/C_5^A(Q^2)$
 should be 
described by the relations given in Eq.~(\ref{pole dominance})
with no adjustable parameters.
In Figs.~\ref{fig:HAoverGA} and \ref{fig:CA6overCA5} 
we show, with  the dashed lines,
the resulting curves for the case of  quenched
lattice results at the lightest pion mass obtained assuming
 the relations
given in Eqs.~(\ref{pole dominance}). As can be seen, as  $Q^2\rightarrow 0$,
both ratios increase slower than pion pole dominance predicts.
In addition we show 
by the solid curves  fits to the same  quenched data
using a dipole form
\be
 \frac{c_0}{(Q^2/m^2+1)}
\ee
 with fit parameters  $c_0$ and $m$. 
The values of $c_0$ and $m$ extracted from the fits are given in 
Table~III. 
In the quenched case, we find that
$m>m_\pi$, whereas for  the hybrid scheme, although $m\sim m_\pi$,
$c_0$ is smaller than $m_N^2/m_\pi^2$ 
causing the dotted line shown in Fig.~\ref{fig:CA6overCA5}, obtained
at the lightest quark mass,
to be higher than the corresponding lattice results. 
The calculation in the hybrid
scheme at the lightest quark mass is done on a larger lattice
enabling us to compute the form factors at low $Q^2$-values,
much lower than in the case of dynamical Wilson fermions.
These results show clear deviations from quenched results at
low $Q^2$, where pion clouds effects are expected to dominate.

\begin{widetext}
\begin{center}
\begin{table}[h]
\begin{tabular}{cccccccccc}
 \hline \multicolumn{10}{c}
{Nucleon elastic}\\
\hline  $m_\pi$~(GeV) & $m$ (GeV) & $c_0$ & $m_A$ (GeV) & $g_0$ & $\tilde{m}_A$ (GeV) & $\tilde{g}_0$ & $\Delta$ &\multicolumn{2}{c}{$g_{\pi NN}$} \\ 
\hline 
\multicolumn{10}{c}
{Quenched Wilson fermions}\\
\hline
       0.563(4) & 0.671(14)& 13.71(34) & 1.659(20) & 1.088(8) & 1.271(9) & 1.074(5) & 0.110(2) & 9.943(99) & 10.609(73)\\
       0.490(4) & 0.597(14)& 15.23(43) & 1.632(19) & 1.079(7) & 1.249(9) & 1.069(5) & 0.083(2) & 9.126(93) & 10.143(91)\\
       0.411(4) & 0.511(16)& 17.70(76) & 1.578(28) & 1.080(12)& 1.220(10)& 1.066(6) & 0.062(2)& 8.410(100)& 9.725(140)\\
\hline
\multicolumn{10}{c}
{$N_F=2$ dynamical Wilson fermions}\\
\hline
0.691(8) &  0.750(43)  & 14.13(1.01) &1.831(22) &  1.067(6)  & 1.393(16) & 1.063(6) & 0.114(3)  & 11.48(245) &10.486(122)\\
0.509(8) &             &             &1.709(46) &  0.999(17) & 1.296(29) & 0.995(17)& 0.038(15) &          & 9.071(294)\\
0.384(8) &  0.642(77)  & 11.15(1.82) &2.019(78) &  0.951(18) & 1.528(44) & 0.943(15)& 0.044(10) &          & 8.613(551)\\
\hline  \multicolumn{10}{c}
{Nucleon to $\Delta$}\\
\hline  $m_\pi$~(GeV) & $m$ (GeV) & $c_0$ & $m_A$ (GeV) & $g_0$ & $\tilde{m}_A$ (GeV)&  $\tilde{g}_0$ &  $\Delta^\prime$ &\multicolumn{2}{c}{$g_{\pi N\Delta}$} \\ 
\hline 
\multicolumn{10}{c}
{Wilson fermions quenched}\\
\hline 
 0.563(4) & 0.691(13) & 3.44(79)& 1.544(32) & 0.952(16) & 1.205(8)  &  0.926(5) & 0.106(2) & 16.560(194) & 17.174(166)\\
 0.490(4) & 0.624(15) & 3.75(12)& 1.537(33) & 0.930(14) & 1.192(10) &  0.910(7) & 0.079(2) & 14.692(188) & 16.195(206) \\
 0.411(4) & 0.545(16) & 4.23(17)& 1.534(36) & 0.906(15) & 1.189(13) &  0.887(9) & 0.052(2) & 12.609(180) & 14.873(264) \\
\hline
\hline
\multicolumn{10}{c}
{Wilson fermions, dynamical $N_F=2$}\\
\hline
0.691(8) &  0.604(95)  & 4.75(1.04) & 1.696(51) &   0.988(24) & 1.368(13) & 0.937(5)  & 0.109(2) &  & 17.536(190) \\
0.509(8) &  0.352(151) & 8.38(6.11) & 1.760(59) &   0.865(25) & 1.454(46) & 0.808(20) & 0.063(2) &  & 14.970(452)\\
0.384(8) &  0.379(58)  & 6.34(1.69) & 1.968(118)&   0.843(40) & 1.410(51) & 0.808(24) & 0.024(15)&  & 12.685(1.416)\\
\hline
\multicolumn{10}{c}
{Hybrid action}\\
0.594(1) & 0.576(28) & 5.08(22)   & 1.924(85)  & 0.883(22) & 1.477(42) &  0.868(15) & 0.076(5)& & 17.649(236)\\
0.498(3) & 0.485(27) & 6.15(52)   & 1.892(101) & 0.864(32) & 1.505(71) &  0.835(27) &0.0648(7) & &17.329(496)\\ 
0.357(2) & 0.389(18) & 8.59(64)   & 1.849(71)  & 0.760(18) & 1.522(72) &  0.708(19) & 0.054(5) &10.053(324)& 14.815(558)\\
\hline
\end{tabular}
\label{Table:fit params}
\caption{The first column gives the pion mass in GeV, the second and third
columns the fit parameters $m$ and $c_0$ extracted from  fitting the ratio
$G_p/G_A$ ($C_6^A/C_5^A$) for the nucleon (N to $\Delta$) case, the
fourth and fifth columns the dipole parameters $m_A$ and $g_0$ extracted from
fitting $G_A$ ($C_5^A$) for the nucleon (N to $\Delta$) and the sixth
and seventh  columns the corresponding parameters but using an exponential
Ansatz $\tilde{g}_0 \exp(-Q^2/\tilde{m}_A^2)$. 
The eight column gives $\Delta$ or $\Delta^\prime$ 
defined in Eq.~(\ref{Gpi linear}).
The last two columns give
the value of the strong coupling constants 
$g_{\pi NN}$ or $g_{\pi N\Delta}$. The first value
of the strong coupling constant is determined using
the fit function of Eq.~(\ref{fit G}),
 whereas the second using a linear fit according 
to Eq.~(\ref{Gpi linear}). }
\end{table}
\end{center}

\end{widetext}

\begin{table}[h]
\begin{tabular}{cccc}
 \hline 
$\kappa$ or $m_l$&  $am_q$ & $af_\pi/Z_A$ &  $Z_A$  \\ 
\hline
\multicolumn{4}{c}{Quenched Wilson fermions} \\
\hline
0.1554 &  0.0403(4) & 0.0611(14) & 0.808(7)~\cite{ZA} \\
0.1558 &  0.0307(4) & 0.0587(16) & 0.808(7)  \\
0.1562 &  0.0213(4) & 0.0563(17) & 0.808(7)  \\
\multicolumn{4}{c}{$N_F=2$ Wilson fermions} \\
\hline
0.1575  &  0.0441(4) & 0.0649(8) & 0.77(2)~\cite{ZA dyn}  \\
0.1580  &  0.0229(4) & 0.0494(9) & 0.78(4)~\cite{ZA dyn}   \\
0.15825 &  0.0122(3) & 0.0467(13)& 0.8\footnote{Estimated from the values
of $Z_A$ at $\kappa=0.1575$ and $0.1580$}
   \\
\hline
\multicolumn{4}{c}{Hybrid action} \\
0.03  &  0.0475(3) & 0.0678(6)   & 1.1085(5)~\cite{LHPC_axial}       \\
0.02  &  0.0324(4) & 0.0648(8)   & 1.0994(4)~\cite{LHPC_axial}  \\
0.01  &  0.0159(2) & 0.0636(6)   & 1.0847(6)~\cite{LHPC_axial}      \\ 
\hline
\end{tabular}
\label{Table:fpi and mq}
\caption{The first column gives the hopping parameter $\kappa$ for
Wilson fermions or the mass of the domain wall fermion, 
the second the renormalized quark mass, 
the third the unrenormalized pion decay constant $f_\pi/Z_A$ in lattice units, 
 and
the fourth the axial renormalization constant $Z_A$.}
\end{table}

In order to examine the $Q^2$-dependence of the
form factors  separately and
 compare with continuum quantities, we need to multiply lattice
results with the axial renormalization constant $Z_A$. These constants
are  known for both Wilson fermions and DWF
within the hybrid scheme. The values that
we use are given in Table~IV. 
We collect our lattice results for 
the  nucleon form factors in 
Tables~V and VI 
 and for $N$ to $\Delta$ in Tables~VII, VIII and IX  
of the Appendix. 
All errors are calculated
using jackknife analysis. 
In Fig.~\ref{fig:GAHA}, we show
 $G_A(Q^2)$ 
and
the induced pseudoscalar form factor 
$G_p(Q^2)$.
For comparison we
also show  results obtained in the hybrid approach 
from Ref.~\cite{LHPC07}. 
The main observation is again that at the smallest
domain wall quark mass, the hybrid results show deviations. In particular, we
note that the value for the  nucleon axial charge $g_A$
 becomes larger in the hybrid scheme
approaching the experimental value. This is in agreement with 
the findings of Ref.~\cite{LHPC_axial}.
Since there are recent state-of-the-art lattice
studies of $g_A$~\cite{LHPC_axial,QCDSF_axial} we will not discuss
it further here but rather investigate  
the $Q^2$ dependence of the form factors.
We also find that $G_p(Q^2)$ increases more rapidly at low $Q^2$
in the hybrid scheme when the pion mass decreases to about 350~MeV.
 In Fig.~\ref{fig:CA5CA6}
we show the corresponding
N to $\Delta$ transition form factors $C_5^A(Q^2)$ and $C_6^A(Q^2)$.
The hybrid results show the same behavior as 
in the case of  the nucleon form factors, yielding
a different behavior at low $Q^2$ when the pion mass becomes about 350~MeV.
The $Q^2$-dependence of both $G_A(Q^2)$ and $C_5^A(Q^2)$ can be well described
by a dipole Ansatz 
\be
\frac{g_0}{(Q^2/m_A^2+1)^2}
\label{dipole}
\quad.
\ee
This is what is usually 
used to described experimental data for $G_A(Q^2)$
 where a value of  $m_A\sim 1.1$~GeV
is extracted for the axial mass. The same dipole Ansatz is also
used to described $C_5^A(Q^2)$, where an axial mass of 
 $1.28\pm 0.10$~GeV~\cite{Kitagaki} has been found.  
In addition, we fit to an exponential form
given by  $\tilde{g}_0 e^{-Q^2/\tilde{m}_A^2}$. 
Both Ans\"atze describe well our results as can be
seen in Figs.~\ref{fig:GAHA} and \ref{fig:CA5CA6}  where
the two lines, which are fits to quenched lattice results at the smallest 
quark mass, can hardly be distinguished.
 The values of the axial masses extracted from these fits are
given in Table~III. 
We find an axial mass that is larger than what is deduced from experiment.
This  means that $G_A(Q^2)$ and $C_5^A(Q^2)$ fall
 off slower than in experiment.
This is clearly seen in Fig.~\ref{fig:GAHA} where we include the dipole
curve taking $m_A=1.1$~GeV.
Having fitted $G_A(Q^2)$ and $C_5^A(Q^2)$, 
the $Q^2$-dependence for the
 form factors $G_p(Q^2)$ and $C_6^A(Q^2)$ can be obtained using 
Eq.~(\ref{pole dominance}). The resulting curves are shown by the dashed line
in Figs.~\ref{fig:GAHA} and \ref{fig:CA5CA6} 
 and show deviations at low $Q^2$. In addition we 
show curves that correspond to 
\be
\frac{g_0 c_0}{(Q^2/m_A^2+1)^2(Q^2/m^2+1)} 
\label{fit functions}
\ee
 with $m$ extracted from fitting the ratio of $G_p(Q^2)/G_A(Q^2)$ 
in the case of the nucleon system and $C_6^A(Q^2)/C_5^A(Q^2)$ for 
the N to $\Delta$.
As expected this provides a
good description of the $Q^2$-dependence for both $G_p(Q^2)$ and $C_6^A(Q^2)$
shown by the solid lines, which correspond to the
parameters of  the quenched data at
 $\kappa=0.1562$.

\begin{figure}[h]
\epsfxsize=8truecm \epsfysize=5truecm 
\mbox{\epsfbox{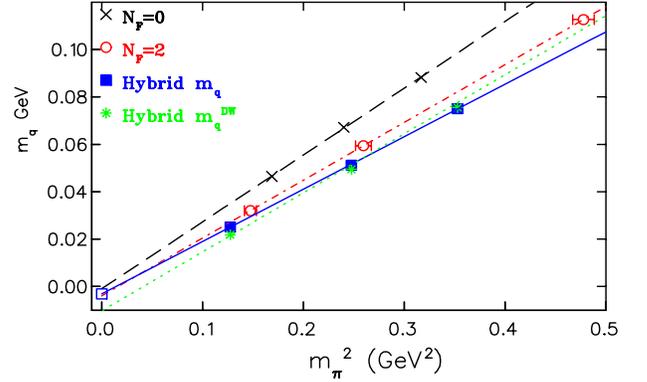}}
\caption{The renormalized quark mass $m_q$
 versus $m_\pi^2$ for the quenched
theory (crosses),  for two dynamical Wilson fermions (open circles) 
and for the hybrid scheme (filled squares).
In the hybrid case we also show $m_q^{\rm DW}$ (asterisks)
 determined by tuning the 
pion mass~\cite{renner}.
The lines are linear fits to $m_\pi^2$. The open square shows
the extrapolated value of  $m_q$ in 
the hybrid scheme. }
\label{fig:quark mass}
\end{figure}

\begin{figure}[h]
\epsfxsize=8truecm \epsfysize=8truecm 
\mbox{\epsfbox{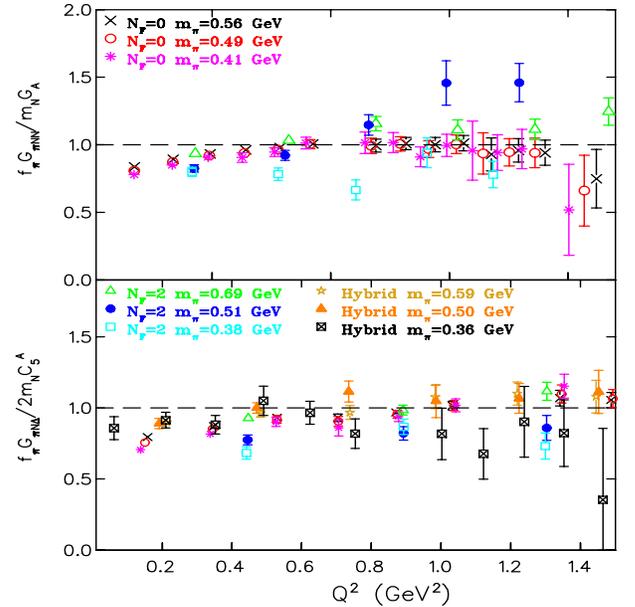}}
\caption{The upper graph shows the ratio $f_\pi G_{\pi NN}(Q^2)/m_NG_A(Q^2)$
and the lower graph the ratio $f_\pi G_{\pi N\Delta}(Q^2)/2m_NC_5^A(Q^2)$. 
The notation is the same as in Fig.~\ref{fig:HAoverGA}. }
\label{fig:GTR simple}
\end{figure}

\begin{figure}[h]
\epsfxsize=8truecm \epsfysize=8truecm 
\mbox{\epsfbox{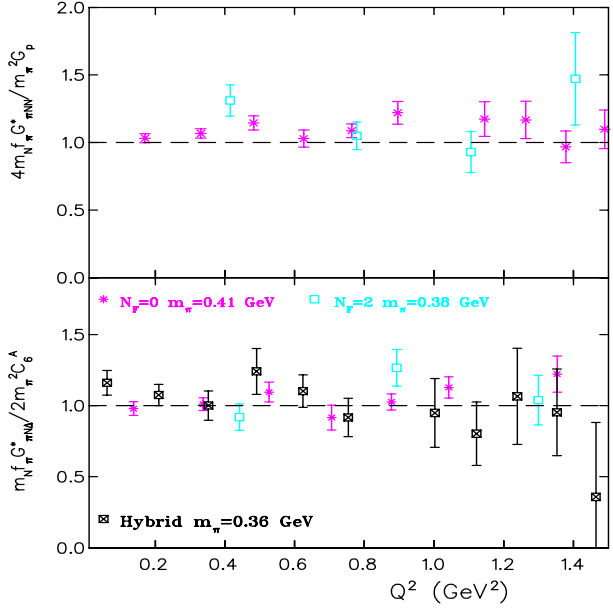}}
\caption{The upper graph shows the ratio 
$4m_Nf_\pi G^*_{\pi NN}(Q^2)/m_\pi^2G_p(Q^2)$
and the lower graph the ratio 
$m_Nf_\pi G^*_{\pi N\Delta}(Q^2)/2m_\pi^2C_6^A(Q^2)$ for the lightest
quark mass considered in each  of our three types of calculations.
We have defined 
$ G^*_{\pi NN}(Q^2)\equiv G_{\pi NN}(Q^2)/(1+Q^2/m_\pi^2)$
with a corresponding expression for $G^*_{\pi N\Delta}$. }
\label{fig:GTR simple2}
\end{figure}

\begin{figure}[h]
\epsfxsize=8truecm \epsfysize=8truecm
\mbox{\epsfbox{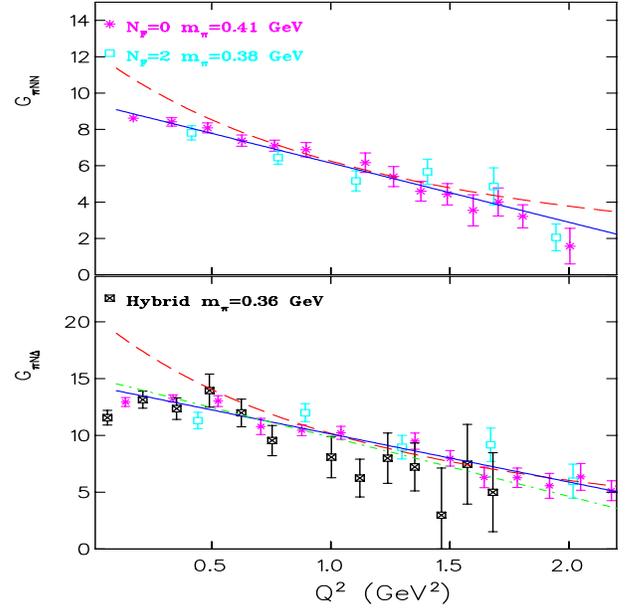}}
\caption{The upper graph shows $G_{\pi NN}(Q^2)$ for Wilson fermions 
at the smallest 
pion mass. The lower graph shows $G_{\pi N\Delta}(Q^2)$
 for Wilson fermions and DWF
at the smallest 
pion mass. The dashed lines follow from the GTRs relations
given in Eq.~(\ref{GTR simple}). The solid lines are fits 
using Eq.~(\ref{Gpi linear}). In the case of $G_{\pi N\Delta}(Q^2)$, we also
show by the dashed-dotted line (with larger slope) the curve that 
corresponds to taking $G_{\pi N\Delta}(Q^2)=1.6\>G_{\pi NN}(Q^2)$.}
\label{fig:Gpi light}
\end{figure}

\begin{figure}[h]
\epsfxsize=8truecm \epsfysize=8truecm 
\mbox{\epsfbox{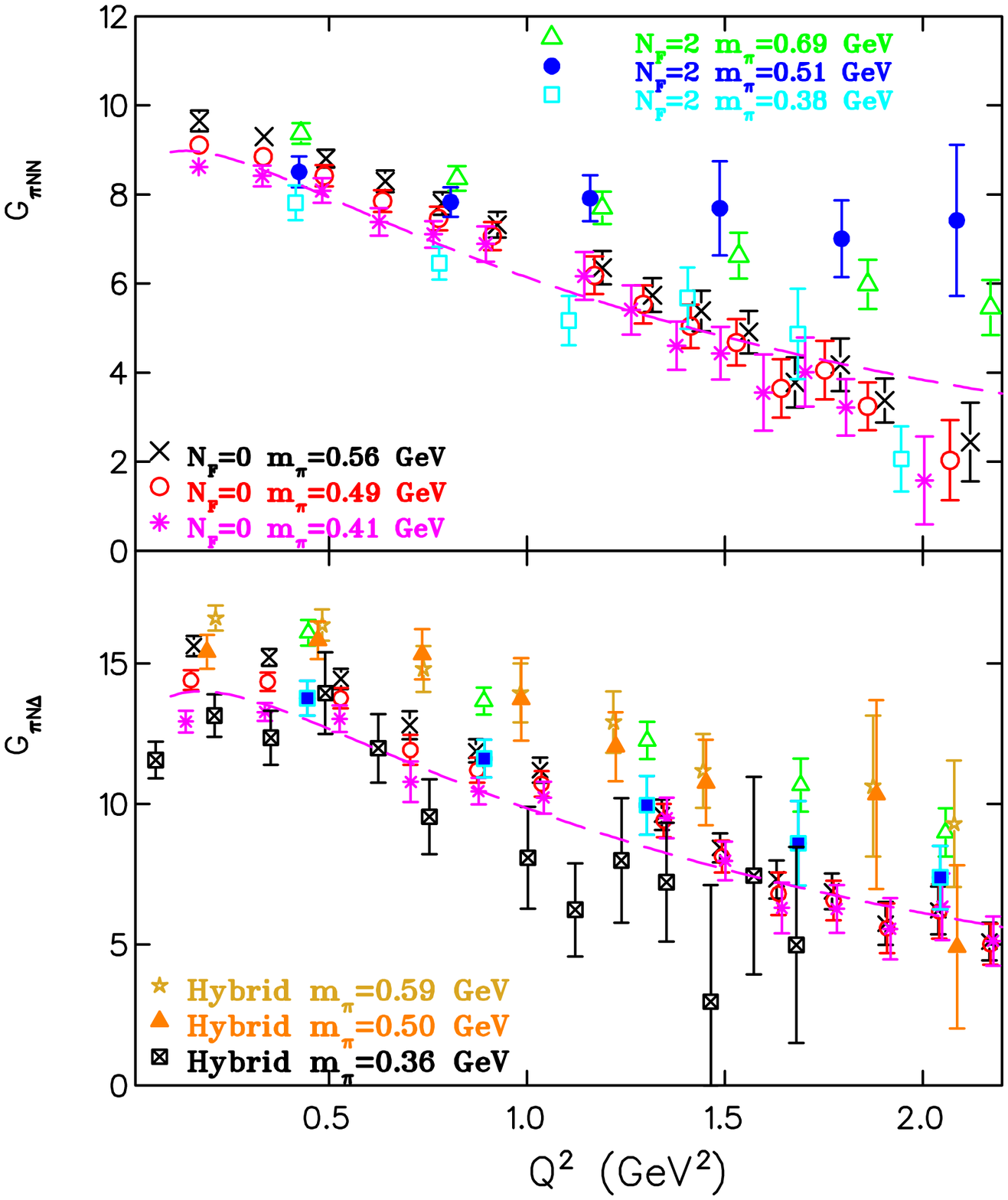}}
\caption{The upper graph shows $G_{\pi NN}(Q^2)$ and  the lower graph shows  
$G_{\pi N\Delta}(Q^2)$ as a function of $Q^2$. 
The dashed lines are fits to the quenched results at $\kappa=0.1562$
obtained using 
the functions of Eq.~\ref{fit functions}
and allowing an overall constant to be fitted.
}
\label{fig:GpiNNandGpiND}
\end{figure}

We now present results that require knowledge of the 
renormalized quark mass.
The renormalized quark mass, $m_q$, is determined by evaluating the
pion to vacuum matrix element of the axial Ward-Takahashi identity
given in Eq.~(\ref{quark mass}). As mentioned already, 
for  Wilson fermions the axial 
Ward identity is satisfied only up  to ${\cal O}(a)$ terms.
We expect these corrections to become more severe as we approach
the chiral limit.
As we already mentioned, the quark mass in the hybrid scheme, $m_q^{\rm DW}$,
was tuned to reproduce the 
mass of the lightest pion in the staggered theory.
Given that domain wall fermions satisfy the AWI
when the size of the fifth dimension is taken to infinity, 
 corrections to  
Eq.~(\ref{quark mass})  come
from the residual mass due the finite length of the fifth direction. 
Therefore differences between the values of $m_q$ and $m_q^{\rm DW}$ are due
to chiral symmetry breaking becuase of the finite size of the fifth dimension.
We show in Fig.~\ref{fig:quark mass}, the renormalized quark mass
extracted from the axial Ward identity for quenched and two dynamical
Wilson fermions and in the hybrid approach. As can be seen, the
pion mass  extrapolates
to zero at $m_q=0$ for the quenched theory.
For  dynamical Wilson
fermions $m_\pi$ is  not exactly zero at $m_q=0$ indicating 
finite $a$-corrections. For the hybrid scheme we show both 
the renormalized mass computed using the AWI, $m_q$, 
and $m_q^{\rm DW}$. As can be seen $m_q$, is
approximately equal to  $m_q^{\rm DW}$ confirming that the residual mass is
small. The biggest deviation, as expected, 
is observe at  the smallest value of the quark mass.
From these results we  confirm that in the hybrid scheme 
 $m_\pi^2\propto m_q$ 
within our statistical errors.

Having determined $m_q$ and $f_\pi$ using Eq.~(\ref{fpi}) we can  
evaluate the form factors
 $G_{\pi NN}(Q^2)$ and   $G_{\pi N\Delta}(Q^2)$. We first examine 
the Goldberger-Treiman relations as given in
 Eq.~(\ref{GTR simple})
by considering the   ratios $f_\pi G_{\pi NN}(Q^2)/m_N G_A(Q^2)$ and
 $f_\pi G_{\pi N\Delta}(Q^2)/2m_N C_5^A(Q^2)$, which should be equal to unity.
 Note
that in these ratios the axial and pseudoscalar renormalization constants
cancel. 
These ratios are shown in Fig.~\ref{fig:GTR simple}.
We find that, in the quenched theory, 
 they are less than one for small $Q^2$
but become one for $Q^2\stackrel{>}{\sim} 0.5$~GeV$^2$.
This is also approximately true for dynamical Wilson for the
two heaviest quark mass.
Results in the hybrid approach, on the other hand, show smaller
deviations from unity at low $Q^2$.
We also expect that the ratios 
\be
\frac{4m_Nf_\pi}{m_\pi^2 G_p(Q^2)}\>\frac{G_{\pi NN}(Q^2)}{(1+Q^2/m_\pi^2)} \quad,\hspace*{0.3cm}
\frac{m_N f_\pi}{2m_\pi^2 C_6^A(Q^2)}\>\frac{G_{\pi N\Delta}(Q^2)}{(1+Q^2/m_\pi^2)} ,
\label{check pole}
\ee
should be unity if pion pole dominance is valid.
 As can be seen in Fig.~\ref{fig:GTR simple2},
for the lightest quark masses in our three types
of action these ratios are indeed consistent with one.

Finally we discuss the $Q^2$-dependence of
 the form factors $G_{\pi NN}(Q^2)$ and 
 $G_{\pi N\Delta}(Q^2)$ separately. In Fig.~\ref{fig:Gpi light} we show 
$G_{\pi NN}(Q^2)$ and $G_{\pi N\Delta}(Q^2)$ at the
smallest quark mass in the quenched theory and for dynamical Wilson
quarks. For  $G_{\pi N\Delta}(Q^2)$ we also
show results in the hybrid scheme at a similar quark mass.
Results for these form factors at small $Q^2$ 
are consistent with each other unlike the other form factors,
indicating that
unquenching effects on these quantities are small.  
Assuming PCAC and pion pole dominance, 
the $Q^2$-dependence  of $G_{\pi NN}(Q^2)$  and $G_{\pi N\Delta}(Q^2)$ 
is completely determined from  the GTRs given in Eq.~(\ref{GTR simple})
once we know $G_A(Q^2)$ and $C_5^A(Q^2)$.
Using the dipole Ansatz of Eq.~(\ref{dipole}) for the $Q^2$-dependence
of $G_A(Q^2)$ and $C_5^A(Q^2)$ with the parameters given in Table~III,
we obtain the dashed 
lines shown  in  Fig.~\ref{fig:Gpi light}. The discrepancy 
already observed in the 
ratio  $f_\pi G_{\pi NN}(Q^2)/m_N G_A(Q^2)$
and $f_\pi G_{\pi N\Delta}(Q^2)/m_N C_5^A(Q^2)$
at low $Q^2$ values is clearly seen here. 
 Results in the hybrid approach confirm 
deviations from the GTRs at low $Q^2$.  
The $Q^2$-dependence of   $G_{\pi NN}(Q^2)$ and   $G_{\pi N\Delta}(Q^2)$
can be described using a linear Ansatz given by
\beq 
G_{\pi NN}(Q^2)&=&a\biggl(1 -\Delta\frac{Q^2}{m_\pi^2}\biggr),\nonumber\\ 
G_{\pi N\Delta}(Q^2)&=&a^\prime\biggl(1 -\Delta^\prime\frac{Q^2}{m_\pi^2}\biggr)
\label{Gpi linear}
\eeq
with $a$ ($a^\prime$) and $\Delta$ ($\Delta^\prime$) fit parameters.
These linear fits are shown by the solid curves in Fig.~\ref{fig:Gpi light}
and provide a good description to the results. Note that we have excluded
from the fits the value at the lowest $Q^2$ in all cases in the quenched theory
as well as in the hybrid approach at the smallest quark mass, since our
statistical error maybe underestimated for this value of $Q^2$.
The values we  find for the parameters $a$ ($a^\prime$), which determine 
 the strong coupling constant $g_{\pi NN}$ ($g_{\pi N\Delta}$),
 and $\Delta$ ($\Delta^\prime$) are
given in Table~III. 
Note that  
$\Delta$ and $\Delta^\prime$ decrease as the quark mass decreases.
In the quenched theory at the smallest quark mass $\Delta \sim 6\%$.
One expects that this value decreases further as we approach the physical
limit becoming comparable to the value of  $\Delta=2.44\%$
obtained
using baryon  chiral perturbation theory~\cite{HBCHT}.
However, the corresponding value of $a$ is smaller than  the expected value
of  $m_N g_A/f_\pi$ and decreases with the quark mass
being  
about 83\% less at the heaviest and about 73\%
 at the lightest quark mass in the quenched theory. 
 The relation $G_{\pi N\Delta}(Q^2)=1.6G_{\pi N N}(Q^2)$
can be used to determine the $Q^2$ behavior of  
 $G_{\pi N\Delta}$ from that of $G_{\pi NN}$.
The dashed-dotted line
in Fig.~\ref{fig:Gpi light}  shows the resulting curve, which approximates
the results well. 

In Fig.~\ref{fig:GpiNNandGpiND}, we  show
results for $G_{\pi N\Delta}(Q^2)$ and $G_{\pi N N}(Q^2)$
 at all values of the quark masses considered in this work.
Assuming pion pole dominance
we relate the $G_{\pi NN}$ and $G_{\pi N\Delta}$ to
$G_P$ and $C_6^A$ respectively using
Eq.~(\ref{pole dominance}). Taking the functions defined in Eq.~(\ref{fit functions}) for the $Q^2$-dependence of $G_p(Q^2)$ and
$C_6^A(Q^2)$ , 
we write 
\be
G_{\pi NN}(Q^2)=K_N\>\frac{Q^2/m_\pi^2+1}{(Q^2/m_A^2+1)^2(Q^2/m^2+1)} 
\label{fit G}
\ee
with a corresponding expression for $G_{\pi N\Delta}(Q^2)$.
The only free parameter is  an overall constant, $K_N$, to be fitted to the
results. The value of $K_N$ determines $g_{\pi NN}$ and
is given in Table~III. The fits to the quenched data at the smallest
quark mass using Eq.~(\ref{fit G}) are shown
by the dashed lines  in Fig.~\ref{fig:GpiNNandGpiND}.
Note that if $m$ were the pion mass, 
then the pole term would cancel, leaving a dipole $Q^2$-dependence
for these form factors.
Allowing $m\ne m_\pi$ and adjusting the overall strength,
we can obtain a reasonable description
of the $Q^2$- dependence of $G_{\pi NN}$
and $G_{\pi N\Delta}$.
 Since lattice results for these form factors 
 do not increase at low values of $Q^2$ as fast as 
expeced by PCAC,
we obtain
a smaller value at $Q^2=0$. 
The values of
$g_{\pi NN}=G_{\pi NN}(0)$
and $g_{\pi N\Delta}=G_{\pi N\Delta}(0)$ extracted using Eq.~(\ref{fit G}),
 when the fit yields $\chi^2/{\rm d.o.f.}\stackrel{<}{\sim}1.5$,
 are given in Table~III. 
If one uses the relation
$G_{\pi NN}(Q^2)=(m_N/f_\pi) G_A(Q^2)$  as the GTR would suggest, then
the extrapolated value at the lightest pion mass 
would be $g_{\pi NN}=11.8 \pm 0.3$ in the quenched theory 
closer to  the experimental value
of $13.21^{+0.11}_{-0.05} $~\cite{exper gpiNN}.
Therefore the different low $Q^2$ dependence observed
in the lattice results  compared to what  is
usually assumed, is responsible for the lower values
of $g_{\pi NN}$
and $g_{\pi N\Delta}$ extracted from these fits.

\section{Summary and Conclusions}
We have presented results for the nucleon axial-vector form factors 
$G_A(Q^2)$ and
$G_p(Q^2)$ as well as for the corresponding 
N to $\Delta$ axial transition form factors $C_5^A(Q^2)$ and $C_6^A(Q^2)$. 
The $\pi NN$
and $\pi N\Delta$ form factors $G_{\pi NN}(Q^2)$ and $G_{\pi N \Delta}(Q^2)$
are also evaluated. 
Using
ratios that show very weak quark mass dependence and are therefore 
 expected to have the same value in the physical limit,
we reach a number of  phenomenologically important conclusions.
One of the main conclusions is that  $G_{\pi NN}$ and $G_{\pi N \Delta}$
have  the same  $Q^2$ dependence yielding a ratio of 
$G_{\pi N\Delta}(Q^2)/ G_{\pi NN}(Q^2)= 1.60(2)$ in good agreement
with what is expected phenomenologically. Similarly the ratio 
$2C_5^A(Q^2)/G_A(Q^2)=1.63(1)$
is also independent of $Q^2$. Equality of these two ratios implies the 
Goldberger-Treiman relations. 
The ratio $8C_6^A(Q^2)/G_p(Q^2)$ on the other hand is larger by about 6\%.
The popular pion pole dominance hypothesis is examined using
our lattice results. We find that in the quenched
theory the ratios of $G_p(Q^2)/G_A(Q^2)$ and $C_6^A(Q^2)/C_5^A(Q^2)$
 require a larger pole mass  $m$ than the corresponding pion mass
in order to  get a good description at low values of $Q^2$. 
On the other hand,  these ratios in the hybrid approach
are well described with $m\sim m_\pi$. However the overall strength
differs from what is predicted using the Goldberger-Treiman relations. 

We also studied the $Q^2$-dependence of the form factors separately.
Comparing quenched and unquenched results at
pion mass of about 350~MeV, we observe large unquenching effects
on the low $Q^2$-dependence of the four form factors, $G_A(Q^2), \>G_P(Q^2), 
\>C_5^A(Q^2)$ and $C_6^A(Q^2)$. This
confirms the expectation that pion cloud effects are expected
to be large at low $Q^2$.
We find that the $Q^2$-dependence of the
 form factors $G_A$ and $C_5^A$ is well described by a dipole of the
form $c_0/(1+Q^2/m_A^2)^2$. For the pion masses used
in this work we find that $m_A\stackrel{>}{\sim} 1.5$~GeV 
as compared to $1.1$~GeV used to describe the $Q^2$ of experimental data
for $G_A$. An exponential
Ansatz of the form $\tilde{c}_0 \exp(-Q^2/\tilde{m}_A^2)$
also provides a good description of the $Q^2$-dependence.
In agreement with a recent lattice evaluation of
 the nucleon axial charge $g_A$~\cite{LHPC_axial}, we find
that $g_A$ increases when using unquenched configurations
on a large volume and pion mass of about 350 MeV
becoming consistent with experiment.
A different low $Q^2$-dependence is observed 
for both quenched and unquenched results in the
case of $G_{\pi NN}(Q^2)$ 
and  $G_{\pi N \Delta}(Q^2)$.
 Instead of the dipole form expected from 
the Goldberger-Treiman relations of Eq.~(\ref{GTR simple}) one finds
that  $G_{\pi NN}$ 
and  $G_{\pi N \Delta}$, in the limit $Q^2\rightarrow 0$ 
 increase less rapidly.
As a result of this, the
 values that we extract for the strong coupling constants
  $g_{\pi NN}=\lim_{Q^2\rightarrow 0}G_{\pi NN}(Q^2)$ and  
$g_{\pi N\Delta}=\lim_{Q^2\rightarrow 0}G_{\pi N\Delta}(0)$ are smaller
than  those extracted from experiment.
One ingredient that is needed for the determination of these
form factors is the renormalized quark mass. In this work, we use the
 axial Ward identity to determine it.
On the lattice, the axial Ward identity 
has ${\cal O}(a)$ corrections in the case of
Wilson fermions, which can become important in particular as we approach
the physical limit.
In the case of domain wall fermions, the axial Ward identity
is only  exact in the limit of large  fifth dimension with
corrections due to the residual mass, which again  become more 
important in the chiral limit.
 The renormalized quark mass, however, affects
 the 
overall strength of these form factors and therefore it can not explain
the different $Q^2$-dependence. To
investigate this issue further one would like to use 
lighter quark masses on a finer lattice. This will become feasible in
the near future as such dynamical simulations are under way.   

 {\bf Acknowledgments:}
We would like to thank B. Orth,  Th. Lippert and K. Schilling~\cite{TchiL}
as well as  C. Urbach, K. Jansen, A. Shindler
and U. Wenger~\cite{Carsten} and the MILC collaboration
 for providing the unquenched
configurations used in this work, as well as the LPHC collaboration for
providing forward propagators~\cite{renner}.
A.T. would like to acknowledge support by the University of Cyprus and the program
``Pythagoras'' of the Greek Ministry of Education.
The computations for this work were partly
carried out on the IBM machine at NIC, Julich,
Germany. This work is
supported in part by the  EU Integrated Infrastructure Initiative
Hadron Physics (I3HP) under contract RII3-CT-2004-506078 and by the
U.S. Department of Energy (D.O.E.) Office of Nuclear Physics under contract 
DE-FG02-94ER40818.
This research used resources of the National Energy Research Scientific
Computing Center, which is supported by the Office of Science of the U.S.
Department of Energy under Contract No. DE-AC03-76SF00098 and
 the MIT Blue Gene computer, supported by the DOE under grant DE-FG02-05ER25681.



\begin{table}[h]
\section{Appendix}
\begin{tabular}{cccc}
 \hline \multicolumn{4}{c}
{Nucleon elastic:Quenched Wilson fermions} \\
\hline 
$Q^2$~(GeV$^2$) & $G_A/Z_A$ & $G_p/Z_A$ & $G_{\pi NN}$ \\
\hline \multicolumn{4}{c}{ $m_\pi=0.563(4)$~(GeV)} \\
\hline
   0.0  & 1.332(17) &             & \\
   0.17 & 1.193(13) & 12.117(188) & 9.645(232) \\
   0.34 & 1.077(16) & 8.456(169)  & 9.296(139)\\
   0.49 & 0.977(15) & 6.227(153)  & 8.804(202) \\
   0.64 & 0.893(15) & 5.102(157)  & 8.294(240)\\
   0.79 & 0.825(18) & 4.123(95)   & 7.799(254) \\
   0.93 & 0.752(18) & 3.277(111)  & 7.318(286)  \\
   1.19 & 0.662(25) & 2.502(117)  & 6.352(374)  \\
   1.32 & 0.589(27) & 2.037(102)  & 5.738(380)  \\
   1.44 & 0.557(31) & 1.885(114)  & 5.384(456)  \\
   1.56 & 0.502(32) & 1.563(106)  & 4.910(475)  \\
   1.68 & 0.421(38) & 1.220(141)  & 3.780(565)  \\
   1.79 & 0.451(47) & 1.307(147)  & 4.177(590)  \\
   1.90 & 0.371(40) & 1.001(119)  & 3.374(497)  \\
   2.12 & 0.337(86) & 0.951(274)  & 2.437(881)  \\
\hline \multicolumn{4}{c}{ $m_\pi=0.490(4)$~(GeV)} \\
\hline
  0.0  & 1.325(15) &               & \\ 
  0.17 & 1.183(15) &  12.432(321)  & 9.109(190)\\
  0.33 & 1.065(21) &  8.352(147)   & 8.845(166)\\
  0.49 & 0.961(23) &  5.964(169)   & 8.419(243) \\
  0.64 & 0.879(20) &  4.914(166)   & 7.850(247)  \\
  0.78 & 0.811(15) &  3.902(116)   & 7.457(262) \\
  0.91 & 0.738(22) &  3.044(109)   & 7.065(318) \\
  1.17 & 0.654(28) &  2.317(124)   & 6.183(421) \\
  1.29 & 0.577(30) &  1.874(108)   & 5.534(428) \\
  1.41 & 0.546(33) &  1.750(119)   & 5.037(487) \\
  1.53 & 0.489(36) &  1.426(110)   & 4.677(518) \\
  1.64 & 0.409(42) &  1.094(147)   & 3.644(657) \\
  1.75 & 0.451(55) &  1.234(161)   & 4.055(651)  \\
  1.86 & 0.363(45) &  0.917(126)   & 3.246(537) \\
  2.07 & 0.323(93) &  0.853(282)   & 2.033(903)  \\
\hline \multicolumn{4}{c}{ $m_\pi=0.411(4)$~(GeV)} \\
\hline
  0.0  & 1.319(19) &   & \\
  0.17 & 1.177(19) & 12.824(345) & 8.619(182)\\
  0.33 & 1.054(22) & 8.225(202)  & 8.417(232)\\ 
  0.48 & 0.947(19) & 5.649(185)  & 8.085(277)\\ 
  0.63 & 0.867(28) & 4.700(199)  & 7.380(309)\\  
  0.76 & 0.798(21) & 3.650(117)  & 7.102(301) \\ 
  0.90 & 0.723(27) & 2.762(120)  & 6.886(393)\\ 
  1.14 & 0.647(36) & 2.084(142)  & 6.167(535)\\ 
  1.26 & 0.566(35) & 1.685(117)  & 5.405(550)\\ 
  1.38 & 0.538(40) & 1.601(128)  & 4.601(541)\\
  1.49 & 0.474(41) & 1.269(118)  & 4.434(589)\\  
  1.60 & 0.394(47) & 0.938(164)  & 3.549(854)\\ 
  1.70 & 0.453(67) & 1.148(187)  & 4.009(777)\\
  1.81 & 0.353(52) & 0.817(142)  & 3.217(633) \\
  2.00 & 0.324(115)& 0.778(322)  & 1.578(986) \\
\hline
\end{tabular}
\label{Table:Quenched results nucleon}
\caption{The first column gives the $Q^2$ in GeV$^2$, the second $G_A$, the
third $G_p$ and the fourth $G_{\pi NN}$ for quenched Wilson fermions.
The errors quoted are jackknife errors.}
\end{table}

\begin{center}
\begin{table}[h]
\begin{tabular}{cccc}
 \hline \multicolumn{4}{c}
{Nucleon elastic: $N_F=2$ Wilson fermions} \\
\hline 
$Q^2$~(GeV$^2$) & $G_A/Z_A$ & $G_p/Z_A$ & $G_{\pi NN}$ \\
\hline \multicolumn{4}{c}{  $m_\pi=0.691(8)$~(GeV)} \\
\hline 
0.0  & 1.382(11) &  & \\ 
 0.43 & 1.105(16) & 8.976(298) & 9.367(228) \\
 0.82 & 0.893(16) & 5.115(170) & 8.361(278) \\
 1.19 & 0.735(22) & 3.075(172) & 7.696(361) \\
 1.53 & 0.658(36) & 2.575(193) & 6.621(516) \\
 1.86 & 0.592(44) & 2.033(190) & 5.980(552) \\ 
 2.17 & 0.484(45) & 1.325(159) & 5.460(617) \\
 2.75 & 0.356(91) & 1.108(322) & 2.441(788) \\
 3.03 & 0.290(74) & 0.678(212) & 3.081(853) \\
\hline \multicolumn{4}{c} {  $m_\pi=0.509(8)$~(GeV)} \\
\hline
  0.0  & 1.270(27) &  & \\
  0.42 & 0.982(21) & 7.229(388) & 8.505(350)\\
  0.81 & 0.809(24) & 3.914(221) & 7.827(331) \\
  1.16 & 0.657(32) & 1.983(225) & 7.913(517)\\
  1.49 & 0.502(50) & 0.668(218) & 7.689(1.057)\\
  1.79 & 0.457(43) & 0.663(151) & 7.005(860) \\
  2.09 & 0.377(72) & 0.300(161) & 7.416(1.691)  \\
  2.63 & 0.216(74) & 0.702(143) & 4.551(1.841)  \\
  2.88 & 0.104(60) & -0.025(84) & 3.298(1.897)    \\
  3.12 & 0.028(25) & 0.226(54)  & 0.651(0.442) \\
 \hline \multicolumn{4}{c} {$m_\pi=0.384(8)$~(GeV)} \\
\hline
  0.0  & 1.175(33)  &  \\ 
  0.42 & 0.990(25)  & 5.418(370)  & 7.812(391)\\
  0.78 & 0.835(32)  & 3.395(244)  & 6.454(366)\\
  1.11 & 0.786(49)  & 2.269(221)  & 5.166(549) \\
  1.41 & 0.611(50)  & 1.270(220)  & 5.673(689) \\
  1.68 & 0.632(98)  & 1.361(223)  & 4.867(1.017) \\
  1.95 & 0.476(77)  & 0.979(177)  & 2.058(734) \\
  2.43 & 0.092(527) & 0.241(1.377)& 0.638(5.365)  \\
 \hline
\end{tabular}
\label{Table:Nf=2 results nucleon}
\caption{The first column gives the $Q^2$ in GeV$^2$, the second $G_A$, the
third $G_p$ and the fourth $G_{\pi NN}$ for Wilson fermions.
The errors quoted are jackknife errors.}
\end{table}
\end{center}

\begin{center}
\begin{table}[h]
\begin{tabular}{cccc}
 \hline \multicolumn{4}{c}
{N to $\Delta$: Quenched Wilson fermions} \\
\hline 
$Q^2$~(GeV$^2$) & $C_5^A/Z_A$ & $C_6^A/Z_A$ & $G_{\pi N\Delta}$ \\
\hline \multicolumn{4}{c}{ $m_\pi=0.563(4)$~(GeV)} \\
\hline
  0.16 & 1.016(14) &   2.675(81) & 15.613(356) \\
  0.35 & 0.902(8)  &   1.778(52) & 15.207(298)  \\
  0.53 & 0.803(15) &   1.269(45) & 14.441(361)   \\
  0.70 & 0.712(14) &   1.012(38) & 12.795(501)   \\
  0.87 & 0.635(17) &   0.774(26) & 11.882(415)  \\
  1.03 & 0.571(18) &   0.607(26) & 11.203(442)   \\
  1.34 & 0.465(21) &   0.418(24) & 9.610(538)   \\
  1.48 & 0.414(23) &   0.348(23) & 8.440(515)   \\
  1.63 & 0.371(25) &   0.305(26) & 7.318(681)  \\
  1.77 & 0.330(25) &   0.247(23) & 6.895(639)   \\
  1.90 & 0.295(31) &   0.204(26) & 5.744(764)    \\
  2.03 & 0.273(32) &   0.188(25) & 6.216(856)   \\
  2.16 & 0.230(28) &   0.145(20) & 5.106(669)  \\
  2.42 & 0.172(48) &   0.116(39) & 3.603(1.527)   \\
\hline \multicolumn{4}{c}{ $m_\pi=0.490(4)$~(GeV)} \\
\hline
  0.16 & 0.999(14) &  2.741(105)  &  14.396(351)\\ 
  0.35 & 0.885(16) &  1.741(52)   &  14.343(325) \\
  0.53 & 0.787(16) &  1.202(46)   &  13.758(354) \\
  0.70 & 0.691(18) &  0.950(44)   &  11.918(535) \\
  0.87 & 0.616(19) &  0.712(30)   &  11.200(441) \\ 
  1.03 & 0.553(22) &  0.552(26)   &  10.717(456) \\
  1.34 & 0.450(26) &  0.377(25)   &  9.412(592) \\
  1.48 & 0.401(26) &  0.313(24)   &  8.129(572) \\
  1.62 & 0.357(28) &  0.274(27)   &  6.807(751) \\ 
  1.76 & 0.315(29) &  0.220(24)   &  6.563(708) \\
  1.90 & 0.288(36) &  0.183(27)   &  5.584(883) \\
  2.03 & 0.262(36) &  0.169(26)   &  6.176(966) \\
  2.15 & 0.219(31) &  0.129(21)   &  5.015(735) \\
  2.40 & 0.150(48) &  0.092(37)   &  3.466(1.749) \\
\hline \multicolumn{4}{c}{ $m_\pi=0.411(4)$~(GeV)} \\
\hline
  0.15 &   0.975(19)  &   2.804(129) & 12.928(392)\\
  0.34 &   0.864(16)  &   1.688(71)  & 13.263(315) \\
  0.53 &   0.769(19)  &   1.114(58)  & 13.023(474) \\
  0.71 &   0.668(19)  &   0.876(52)  & 10.784(719) \\
  0.87 &   0.593(23)  &   0.634(34)  & 10.453(479)  \\
  1.04 &   0.536(25)  &   0.488(27)  & 10.225(563) \\
  1.34 &   0.438(30)  &   0.333(28)  & 9.496(718)  \\
  1.49 &   0.391(31)  &   0.272(25)  & 7.978(684)  \\
  1.63 &   0.347(32)  &   0.240(30)  & 6.303(901)  \\
  1.77 &   0.303(34)  &   0.190(26)  & 6.269(848)  \\
  1.90 &   0.284(44)  &   0.157(29)  & 5.555(1.089)  \\
  2.03 &   0.250(44)  &   0.150(28)  & 6.346(1.179)   \\
  2.15 &   0.211(37)  &   0.113(22)  & 5.127(870)  \\
  2.40 &   0.121(50)  &   0.065(37)  & 3.425(2.354)  \\
\hline
\end{tabular}
\label{Table:Quenched results NDelta}
\caption{The first column gives the $Q^2$ in GeV$^2$, the second $C_5^A$, the
third $C_6^A$ and the fourth $G_{\pi N\Delta}$.
The errors quoted are jackknife errors.}
\end{table}
\end{center}

\begin{center}
\begin{table}[h]
\begin{tabular}{cccc}
 \hline \multicolumn{4}{c}
{N to $\Delta$: $N_F=2$ Wilson fermions} \\
\hline 
$Q^2$~(GeV$^2$) & $C_5^A/Z_A$ & $C_6^A/Z_A$ & $G_{\pi N\Delta}$ \\
\hline \multicolumn{4}{c} {$m_\pi=0.691(8)$~(GeV)} \\
\hline 
  0.449   &   0.956(16) &    2.022(93)  & 16.082(457)\\   
  0.882   &   0.765(19) &    1.080(57)  & 13.653(480)  \\
  1.286   &   0.605(23) &    0.586(41)   & 12.253(656)   \\
  1.666   &   0.500(34) &    0.457(48)   & 10.671(945) \\
  2.025   &   0.444(38) &    0.371(39)   & 8.990(863)  \\
  2.367   &   0.361(37) &    0.240(32)   & 8.210(1.029) \\ 
  3.008   &   0.229(63)&     0.162(49)   & 4.431(1.341)  \\
  3.310   &   0.161(45) &    0.083(37)   & 4.961(1.391)  \\
\hline \multicolumn{4}{c} {$m_\pi=0.509(8)$~(GeV)} \\
\hline
  0.441 & 0.845(23)  &      1.517(111)  &  13.761(622)\\ 
  0.895 & 0.673(23)  &      0.744(55)   &  11.611(668)\\
  1.311 & 0.552(29)  &      0.380(46)   &  9.955(1.049)\\
  1.697 & 0.502(55 ) &      0.240(56)   &  8.599(1.500)\\
  2.059 & 0.382(38)  &      0.161(28)   &  7.389(1.124)\\
  2.401 & 0.305(65)  &      0.083(30)   &  7.213(1.880)\\
  3.035 & 0.144(62)  &      0.039(25)   &  3.656(2.079)\\
\hline \multicolumn{4}{c} {$m_\pi=0.384(8)$~(GeV)} \\
\hline
  0.443  &  0.837(28)  &      1.335(96)  & 11.302(722)\\   
  0.891  &  0.700(25)  &      0.583(46)  & 11.997(792) \\
  1.293  &  0.620(34)  &      0.381(35)  & 8.956(1.0358) \\
  1.661  &  0.530(38)  &      0.256(29)  & 9.168(1.482) \\
  2.002  &  0.452(63)  &      0.203(30)  & 5.966(1.504)  \\
  2.322  &  0.342(48   &      0.136(22)  & 2.054(1.086) \\
  3.183  &  0.102(23)  &      0.030(9)   & 0.950(6.900) \\
\hline
\end{tabular}
\label{Table:Nf=2 results NDelta}
\caption{The first column gives the $Q^2$ in GeV$^2$, the second $C_5^A$, the
third $C_6^A$ and the fourth $G_{\pi N\Delta}$.
The errors quoted are jackknife errors.}
\end{table}
\end{center}

\begin{center}
\begin{table}[h]
\begin{tabular}{cccc}
 \hline \multicolumn{4}{c}
{N to $\Delta$: Hybrid action} \\
\hline 
$Q^2$~(GeV$^2$) & $C_5^A/Z_A$ & $C_6^A/Z_A$ & $G_{\pi N\Delta}$ \\
\hline \multicolumn{4}{c} {  $m_\pi=0.594(1)$~(GeV)} \\
\hline
  0.213  &  0.689(42)  &  2.131(92)  & 16.602(449)  \\    
  0.482  &  0.642(30)  &  1.326(60)  & 16.359(563)   \\
  0.738  &  0.578(28)  &  0.944(57)  & 14.796(808)  \\
  0.983  &  0.490(31)  &  0.614(43)  & 13.944(1.049)  \\
  1.218  &  0.444(30)  &  0.477(32)  & 12.911(1.094) \\
  1.445  &  0.392(32)  &  0.377(31)  & 11.178(1.315)  \\
  1.874  &  0.317(42)  &  0.244(45)  & 10.628(2.508)\\
  2.079  &  0.261(41)  &  0.180(32)  &  9.296(2.253) \\
  2.278  &  0.200(47)  &  0.132(29)  &  9.768(3.075) \\
  2.472  &  0.201(53)  &  0.118(32)  &  10.702(3.698)\\
  2.660  &  0.132(80)  &  0.115(63)  &  4.125(4.734)\\
  2.844  &  0.101(89)  &  0.041(47)  &  11.147(10.192) \\
\hline \multicolumn{4}{c} { $m_\pi=0.498(3)$~(GeV)} \\
\hline
  0.191 & 0.683(33)  &    2.398(186)  &  15.404(608) \\  
  0.471 & 0.621(27)  &    1.212(77)   &  15.823(667)  \\
  0.735 & 0.540(32)  &    0.809(75)   &  15.325(896) \\
  0.985 & 0.548(42)  &    0.686(76)   &  13.718(1.466)\\
  1.224 & 0.446(34)  &    0.445(44)   &  12.040(1.226)\\
  1.452 & 0.385(38)  &    0.332(40)   &  10.760(1.523)\\
  1.882 & 0.296(53)  &    0.191(42)   &  10.334(3.357)\\
  2.087 & 0.284(67)  &    0.182(47)   &  4.914(2.902) \\
  2.284 & 0.325(116) &    0.190(70)   &  2.688(5.406) \\
  2.476 & 0.208(106) &    0.101(59)   &  2.967(5.431)\\
  2.660 & 0.125(134) &    0.102(99)   &   \\
\hline \multicolumn{4}{c} { $m_\pi=0.357(2)$~(GeV)} \\
\hline
  0.059 & 0.560(42) &   3.219(234)  &  11.557(649)\\ 
  0.208 & 0.598(30) &   2.213(159)  &  13.141(758)\\
  0.351 & 0.582(28) &   1.569(131)  &  12.354(956)\\
  0.490 & 0.552(31) &   1.108(100)  &  13.941(1.455)\\
  0.624 & 0.516(30) &   0.882(74)   &  11.980(1.219)\\
  0.754 & 0.484(32) &   0.720(67)   &  9.545(1.330)\\
  1.003 & 0.411(42) &   0.460(67)   &  8.088(1.810) \\
  1.122 & 0.383(41) &   0.379(52)   &  6.238(1.658)\\
  1.239 & 0.368(47) &   0.335(54)   &  7.991(2.222)\\
  1.353 & 0.365(53) &   0.312(57)   &  7.220(2.114)\\
  1.465 & 0.348(79) &   0.318(83)   &  2.968(4.150)\\
  1.574 & 0.315(74) &   0.204(60)   &  7.453(3.508)\\ 
  1.681 & 0.295(89) &   0.196(68)   &  4.989(3.480)\\
  1.888 & 0.194(100)&   0.112(70)   & \\

\hline
\end{tabular}
\label{Table:Hybrid results NDelta}
\caption{The first column gives the $Q^2$ in GeV$^2$, the second $C_5^A$, the
third $C_6^A$ and the fourth $G_{\pi N\Delta}$. 
The errors quoted are jackknife errors.}
\end{table}
\end{center}
\end{document}